\documentclass{article}
\usepackage{arxiv}

\usepackage{rotating}
\usepackage[utf8]{inputenc} % allow utf-8 input
\usepackage[T1]{fontenc}    % use 8-bit T1 fonts
\usepackage{hyperref}       % hyperlinks
\usepackage{url}            % simple URL typesetting
\usepackage{booktabs}       % professional-quality tables
\usepackage{amsfonts}       % blackboard math symbols
\usepackage{nicefrac}       % compact symbols for 1/2, etc.
\usepackage{microtype}      % microtypography
\usepackage{graphicx}
\usepackage[square,numbers]{natbib}
\usepackage{doi}

\usepackage{amsmath, amssymb, amsthm}
\usepackage{enumitem}
\usepackage{lscape}

\theoremstyle{theorem}
\newtheorem{theorem}{Theorem}
\newtheorem{proposition}{Proposition}

\theoremstyle{remark}

\theoremstyle{definition}
\newtheorem{definition}[theorem]{Definition}

\title{\textsf{{A model and method for analyzing the precision of binary measurement methods based on beta-binomial distributions, and related statistical tests}}}

\author{
\href{https://orcid.org/0000-0002-3471-1594}{\includegraphics[scale=0.06]{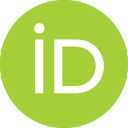}\hspace{1mm}Jun-ichi Takeshita}\thanks{Corresponding author: \texttt{jun-takeshita@aist.go.jp}} \\
        Research Institute of Science for Safety and Sustainability,\\
        National Institute of Advanced Industrial Science and Technology (AIST), Tsukuba, Japan
        \AND
        Tomomichi Suzuki\\
        Department of Industrial Administration, Faculty of Science and Technology\\
        Tokyo University of Science, Noda, Chiba, Japan.
}

\renewcommand{\headeright}{JI Takeshita and T Suzuki}
\renewcommand{\undertitle}{Pre-final version}
\renewcommand{\shorttitle}{} %

\begin{document}
\maketitle

\begin{abstract}
  This study developed a statistical model and method for analyzing the precision of binary measurement methods from collaborative studies.
  The model is based on beta-binomial distributions. 
  In other words, it assumes that the probability of detection (POD) of each laboratory obeys a beta distribution, and the binary measured values under a given POD follow a binomial distribution. 
  We propose the key precision measures of repeatability and reproducibility for the model, and provide their unbiased estimators.
  Then, our precision measures and estimators are theoretically compared to three existing binary methods.
  Further, through consideration of a number of statistical tests for homogeneity of proportions, we propose appropriate methods for determining laboratory effects in the new model. 
  Finally, we apply the results to a real-world example in the field of food safety.
  
\end{abstract}

\section{Introduction}
\label{sec:intro}

In order to promote a new measurement method, it is important to assess not merely the accuracy but the precision of the method. 
To this end, collaborative studies (i.e., studies in which several laboratories measure identical objects, using the same protocol, and compare the measured values) are usually conducted. 
The main aims of collaborative studies are to evaluate the precision of new measurement methods, and to determine whether there is a difference in variability between laboratories. 
The International Organization for Standardization (ISO) 5725 series is widely used in the conduct of such studies and analysis of the results. 
ISO 5725-2~\citep{iso2019ISO572522019Accuracy} assumes that the measured values are based on a population following normal distributions, and uses a one-way analysis of variance (ANOVA) to evaluate the precision; however, several ISO committees and industrial agencies are now interested in assessing binary measurement methods. 
For example, the ISO/ Technical Committee (TC) 34 (Food products) / Subcommittee (SC) 9 (Microbiology) discusses studies on the detection of Listeria monocytogenes in foods~\citep{scotter2001ValidationISOMethod11290} and on new real-time polymerase chain reaction (PCR) assays for detecting transgenic rice~\citep{grohmann2015CollaborativeTrialValidationCry1AbAc}, both of which provide binary measured values.

In the context of the ISO 5725 series, several statistical methods to analyze binary measured values have recently been proposed; among them, \citep{bashkansky2012InterlaboratoryComparisonTestResults,gadrich2012ORDANOVAAnalysisOrdinalVariation,langton2002AnalysingCollaborativeTrialsQualitative, wilrich2010DeterminationPrecisionQualitativeMeasurement}.
\citet{wilrich2010DeterminationPrecisionQualitativeMeasurement} modified the basic model in ISO 5725-2, and directly applied it to evaluate the accuracy of binary measurement methods. 
This model assumes that the binary measured values are based on a population obeying binomial distributions. 
As this assumption is quite reasonable when assessing binary measured values, and the model is easy for users to understand, Wilrich's work is frequently referenced. 
Before that model was developed, \citet{langton2002AnalysingCollaborativeTrialsQualitative} proposed two precision measures, accordance and concordance, to assess the precision of binary measurement methods. 
The two measures are based on the probability that two measured values are identical.

\citet{wilrich2010DeterminationPrecisionQualitativeMeasurement} also conducted statistical tests to detect laboratory effects, by utilizing a chi-squared test for independence in contingency tables. 
However, it is known that chi-squared tests are applicable under two conditions, $np \geq 5$ and $n(1-p) \geq 5$, where $n$ is the number of repetitions in each laboratory and $p$ is a binomial probability; and given these conditions, the number of repetitions $n$ must be more than $10$. 
Wilrich noted this limitation, but proposed no alternative methods. 
However, there are many real-world examples of collaborative studies, especially those dealing with biological responses, that do not fulfill this requirement. 
Such collaborative studies are important in several fields, such as food safety, chemical risk assessment and management, and so on~\citep{iso2016ISO161402016Microbiology,oecd2018TestNo442EVitro}. 
Note that the Organisation for Economic Co-operation and Development (OECD) has a key role in discussing how to conduct chemical risk assessment and management internationally, and the results of collaborative studies provide important information for developing OECD test guidelines. 
However, there are no details on how to conduct statistical analyses in the related OECD guidance documents~\citep{oecd2005No34GuidanceDocument}.

The present study, then, had two aims: first, to propose a new model for analyzing the precision of binary measurement methods, by extending Wilrich's model; and second, to propose appropriate statistical tests to determine whether laboratory effects exist, from the results of collaborative studies. 
The paper is organized as follows. 
Section~\ref{sec:pre} outlines the notation and summarizes the basic relevant facts. 
Section~\ref{sec:bbmodel} presents our new model. 
First, we introduce the basic model, based on beta-binomial distributions, and define precision measures of repeatability and reproducibility for the model. 
Then, we give unbiased estimators of the measures and, using simulated studies, discuss the characteristics of the estimators. 
Section~\ref{sec:comparison} compares the proposed precision measures and their estimators to those of three previous works. 
Section~\ref{sec:test-meth-homog} deals with several statistical tests for homogeneity of proportions, and proposes appropriate methods for the new model, using simulated studies. 
In Section~\ref{sec:ex}, we apply the results to three real-world examples. 
Section~\ref{sec:conc} summarizes the findings of the study.

\section{Preliminaries}
\label{sec:pre}
\subsection{Precision in ISO 5725}

The ISO 5725 series defines the accuracy of measurement methods and results as general terms involved with trueness and precision. 
Trueness, defined as the closeness of agreement between the average value obtained from a large series of measured values and an accepted reference value, is usually expressed in terms of bias. 
In other words, it is the difference between the expectation of the measured values and the accepted reference value. 
Precision, defined as the closeness of agreement between independent measured values obtained under stipulated conditions, is usually expressed in terms of standard deviations of the measured values.

To assess the precision of measurement methods, two measures are typically used: repeatability and reproducibility. 
Repeatability refers to values measured under repeatability conditions; that is, independently measured values obtained by the same method, using identical test objects, in the same laboratory, by the same operator, using the same equipment, within a short time interval. 
Reproducibility relates to reproducibility conditions; that is, measured values are obtained, using the same method, on identical test objects, in different laboratories, with different operators, using different equipment.

In the ISO 5725 series, the basic model for measured values to estimate the accuracy of a given measurement method is as follows:
\begin{equation*}
  y_{ij} = m + B_i + e_{ij},
\end{equation*}
where $y_{ij}$ is the measured value of trial $j$ in laboratory $i$; $m$ is a general mean (expectation); $B_i$ is the laboratory component of variation (under repeatability conditions) in laboratory $i$, whose expectation is assumed to be $0$, and its variance is the between-laboratory variance $\sigma_L^2$; and $e_{ij}$ is a residual error (under repeatability conditions), whose expectation is also assumed to be $0$, and its variance is the within-laboratory variance $\sigma_{ri}^2$.
The present study assumes that $B_i$ and $e_{ij}$ are independent, and the number of repetitions is identical in all the laboratories and it is denoted as $n$.
Moreover, the within-laboratory variance $\sigma_{ri}^2$ is assumed to be identical in all laboratories and is denoted as the repeatability variance $\sigma_r^2$. 
The reproducibility variance $\sigma_R^2$ is defined as $\sigma_R^2{:=}\sigma_r^2+\sigma_L^2$.
Note that the definition of the reproducibility variance in ISO 5725 series differ from that in the Gauge R\&R studies~\citep{burdick2005DesignAnalysisGaugeStudiesa}, but the definition in ISO 5725 series is the first to be defined.

\subsection{Beta-binomial distribution}
\label{ssec:betabinomial}

\begin{definition}
  A random variable $X$ follows a beta-binomial distribution if the probability mass function of the variable $X$ is defined as follows:
  \begin{equation*}
    P(X=x) = \binom{n}{x} \frac{B(x+a, n-x+b)}{B(a, b)}, \quad (x = 0, 1, 2, \ldots, n),
  \end{equation*}
  where $a, b > 0$ are non-negative real numbers, and $B(a, b)$ is a beta function defined as
  \begin{equation*}
    B(a, b) = \int_0^1 p^{a - 1} (1-p)^{b - 1} dp.
  \end{equation*}
\end{definition}
We remark that a beta-binomial distribution is a compound distribution which assumes that the defective ratio parameter, or the binomial probability, $p$, of a given binomial distribution, follows a beta distribution.

If a random variable $X$ follows a beta-binomial distribution, say $BBi (n,a,b)$, then the expectation and variance of $X$ are, respectively:
\begin{equation*}
  E(X) = \frac{n a}{a + b} \quad \text{and} \quad
  V(X) = \frac{n a b ( a + b + n)}{(a + b)^2 (a + b +1)}.
\end{equation*}
Further, let $p:=a / (a+b)$ and $\lambda := 1/(a + b+ 1)$, the expectation and variance can be written as
\begin{equation*}
  E(X) = n p  \quad \text{and} \quad
  V(X) = n p (1-p) [1 + (n-1) \lambda].
\end{equation*}
The $\lambda$ is known as the over-dispersion parameter.

\section{Beta-binomial model}
\label{sec:bbmodel}

This section first introduces a new model using a beta-binomial distribution, and then provides estimators of precision measures based on the model.

\subsection{Our basic model}

We propose the following basic model of the measured value $y_{ij} \in \{0,1\}$, to evaluate the precision of binary measurement methods:
\begin{equation}
  \label{model:proposed}
  \begin{cases}
    p_i          & \sim \text{a beta distribution } \textit{Beta}(a, b),    \\
    y_{ij} | p_i & \sim \text{a Bernoulli distribution } \textit{Be}(p_i),
  \end{cases}
\end{equation}
where $y_{ij}$ is the measured value defined as $0$ or $1$ for trial $j$ in laboratory $i$; and $p_i$ is the probability of detection (POD, i.e., the probability of obtaining a measured value $y_{ij}=1$) for laboratory $i$. 
In other words, if we let $x_i |p_i := \sum_{j=1}^{n} y_{ij}|p_i$ and $x_i := \sum_{j=1}^{n} y_{ij}$, then $x_i|p_i$ and $x_i$ are assumed to follow a binomial distribution $\textit{Bi}(n,p_i)$ and a beta-binomial distribution $\textit{BBi}(n,a,b)$, respectively.

\subsection{Definitions of the repeatability, between-laboratory, and reproducibility variances}

First, the within-laboratory variance of laboratory $i$, say $\sigma^2_{ri}$, is the variance of a random error of trial $j$ in laboratory $i$, and the repeatability variance is the expectation of $\sigma_{ri}^2$. 
Since, for any laboratory $i$, $y_{ij}$ follows a Bernoulli distribution with parameter $p_i$, $\sigma_{ri}^2 := p_i(1-p_i)$ and $\sigma_r^2 := E(p_i (1-p_i))$.
Next, because the between-laboratory variance $\sigma_L^2$ is the variance of a laboratory component of variation, $\sigma_L^2:=Var (p_i)$.
Finally, since the reproducibility variance $\sigma_R^2$ is the sum of the repeatability and between-laboratory variances, $\sigma_R^2 := \sigma_r^2+\sigma_L^2$; and from the definitions of $\sigma_r^2$ and $\sigma_L^2$, we have
\begin{equation}
  \label{eq:8}
  \sigma^2_R = E(p_i) - E(p_i^2) + \mathrm{Var}(p_i) = E(p_i) - E^2(p_i) = E(p_i) \left(1 - E(p_i)\right).
\end{equation}

The definitions of $\sigma_r$, $\sigma_L$, and $\sigma_R$ can be expressed, using the parameters of a beta-binomial distribution (i.e., $a$ and $b$), as follows.
\begin{proposition}
  \label{prop:variances-theoritical}
  
  Assume that $p_i$ and $y_{ij}|p_i$ follow a beta distribution $\textit{Beta}(a,b)$ and a Bernoulli distribution $\textit{Be} (p_i)$, respectively.
  Then the repeatability, between-laboratory, and reproducibility variances are respectively calculated as follows: 
  \begin{align*}
    \sigma^2_r & =  \frac{a b}{(a + b)(a + b + 1)},               \\
    \sigma^2_L & =\frac{a b}{(a + b)^2 (a + b + 1)}, \text{ and } \\
    \sigma^2_R & =\frac{a b}{(a + b)^2}.
  \end{align*}
\end{proposition}

A proof of Proposition~\ref{prop:variances-theoritical} is shown in Appendix~\ref{ssec:pr-variances-theoritical}.
Since the variance of the beta-binomial distribution $\textit{BBi} (n,a,b)$ is $nab (a+b+n) / ((a+b)^2 (a+b+1))$, from Proposition~\ref{prop:variances-theoritical} the proposition below holds, if the number of repetitions in each laboratory $n$ is determined.
\begin{proposition}\label{prop:variances-compare}
  Among the variances $\sigma_r$, $\sigma_L$, $\sigma_R$, and $\sigma_{\textit{BBi}}$, the following relations hold: 
  \begin{align*}
    \sigma^2_L = \frac{\sigma^2_{\textit{BBi}} - n \sigma^2_r}{n^2} \text{ and }
    \sigma^2_R = \frac{\sigma^2_{\textit{BBi}} + n(n-1) \sigma^2_r}{n^2},
  \end{align*}
  where $\sigma_{\textit{BBi}}^2$ is the variance of a given beta-binomial distribution.
\end{proposition}

A proof of Proposition~\ref{prop:variances-compare} is shown in Appendix~\ref{ssec:pr-variances-compare}.

\subsection{Estimators of the repeatability, between-laboratory, and reproducibility variances}

When the numbers of laboratories and repetitions in a collaborative study are determined, we obtain unbiased estimators of the POD and the three precision variances using Proposition~\ref{prop:variances-compare}.
\begin{proposition}\label{prop:est}
  If the number of laboratories, denoted by $L$, and that of repetitions, denoted by $n$, in a collaborative study are determined, then the following are unbiased estimators of $p_i$, $\sigma^2_r$, $\sigma^2_L$, and $\sigma^2_R$, respectively:
  \begin{align}
    \hat{p}_i        & = \frac{1}{n} \sum_{j=1}^n y_{ij},\label{eq:unbiased_est_pi}                                               \\
    %\hat{p} & = \frac{1}{L} \sum_{i=1}^L \hat{p}_i,\label{eq:unbiased_est_p}\\
    \hat{\sigma}_r^2 & = \frac{n \sum_{i=1}^L \hat{p}_i (1 - \hat{p_i})}{L(n-1)},\label{eq:unbiased_est_sigmar}                   \\
    \hat{\sigma}_L^2 & = \frac{\hat{\sigma}_{\textit{BBi}}^2 - n \hat\sigma_r^2}{n^2}, \text{ and }\label{eq:unbiased_est_sigmaL} \\
    \hat{\sigma}_R^2 & = \frac{\hat{\sigma}_{\textit{BBi}}^2 + n(n-1) \hat\sigma_r^2}{n^2},
    \label{eq:unbiased_est_sigmaR}
  \end{align}
  where $\hat{\sigma}_{\textit{BBi}}$ is an estimator of the variance of a given beta-binomial distribution; that is:
  \begin{equation*}
    \hat\sigma^2_{\textit{BBi}} =
    \begin{cases}
      \displaystyle\frac{n^2}{L-1} \sum_{i=1}^L \left( \hat{p}_i - \frac{1}{L} \sum_{i=1}^L \hat{p}_i\right)^2, & 
      \text{ if the expectation of POD is unknown,}                                                                 \\
      \displaystyle\frac{n^2}{L} \sum_{i=1}^L \left( \hat{p}_i - E(p_i)\right)^2,                               & 
      \text{ if the expectation of POD is known.}
    \end{cases}
  \end{equation*}
\end{proposition}

A proof of Proposition~\ref{prop:est} is shown in Appendix~\ref{sec:proof-est}. 

From Proposition~\ref{prop:est}, the estimators~\eqref{eq:unbiased_est_sigmar}, \eqref{eq:unbiased_est_sigmaL}, and \eqref{eq:unbiased_est_sigmaR} have a good property in being unbiased; however, the estimate of the between-laboratory variance can be negative as it is defined using a difference.
In the next subsection, we discuss the properties of the estimators using simulated studies.

\subsection{Simulated studies using the beta-binomial model}
\label{sec:sim_study}

To discuss the properties of the estimators, we simulated collaborative studies in which several beta-binomial models are assumed. 
The following cases were treated: $(a,b)=(13.3,5.7)$, $(6.3,2.7)$, $(0.7,0.3)$, $(17.1,1.9)$, $(8.1,0.9)$, $(0.9,0.1)$, $(18.05,0.95)$, $(8.55,0.45)$, and $(0.95,0.05)$. 
The expectations of $\textit{Beta} (a,b)$ [i.e., $p:=a/(a+b)$] in the first, second, and last three cases are $0.7$, $0.9$, and $0.95$, respectively, and the over-dispersion parameters of $\textit{BBi}(n,a,b)$ [i.e., $\lambda := 1/(a+b+1)$] of each group are $0.05$, $0.1$, and $0.5$, respectively. 
For each case, the number of laboratories $L$ was set at $5$ and $10$, and the number of measured values in each laboratory $n$ was set at $5$, $10$, and $100$. 
Each of these collaborative studies was repeated $10,000$ times.

First, we observe that the average estimates $\hat{p}$, $\hat{\sigma}_r^2$, $\hat{\sigma}_L^2$, and $\hat{\sigma}_R^2$ are almost identical to their theoretical values $p$, $\sigma_r^2$, $\sigma_L^2$, and $\sigma_R^2$, respectively. 
The unbiasedness of the estimators shown in Proposition~\ref{prop:est} was also confirmed by the results of the simulated studies. 
Moreover, if $n$ is larger under the same parameters in $\textit{Beta}(a,b)$, then the ranges covering $95\%$ of the simulated study results of the estimates  $\hat{p}_u- \hat{p}_l$ and $\hat{\sigma}_{\star,u}^2 - \hat{\sigma}_{\star,l}^2 \  (\star\in \{r,L,R\})$ are smaller in all cases, where the subscripts $u$ and $l$ mean the upper and lower $2.5\%$-percentile results.
Comparing the cases where $L=5$ and $10$, where all the parameters except $L$ are identical, we find that the ranges covering $95\%$ of the simulated study results of the estimates for $L=10$ are smaller than for $L=5$ in all cases. 
These facts are obvious and well-known as the law of large numbers.
The average, and the lower and upper $2.5$-percentile results, of $10,000$ simulated collaborative studies, when $L=5$ are shown in Table~\ref{tab:estres_L5}.
The same information as in Table~\ref{tab:estres_L5}, but with $L=10$, are shown in Table~\ref{tab:estres_L10}.

Next, we focus on the ranges covering $95\%$ of the simulated study results for the estimates with the same $L$ and $n$. 
For $L=5$, Table~\ref{tab:range_CI_L5} summarizes the respective ranges of the POD, and the repeatability, between-laboratory, and reproducibility variances, for each $p$, $\lambda$, and $n$. 
Table~\ref{tab:range_CI_L10} shows the same information as in Table~\ref{tab:range_CI_L5}, but with $L=10$. 
In the tables, bold and italic numbers stand for the minimum and maximum values, respectively, in the blocks with the same $n$, and the same ranges of $p$ or precision variances, expressed as $3\times 3$ matrices. 
Hereafter, we call each $3 \times 3$ matrix a block.

\begin{sidewaystable}[htbp]
  \centering
  %  \small
  \caption{Ranges covering $95\%$ simulated experiment results of $p$, $\sigma_r^2$, $\sigma_L^2$, and $\sigma_R^2$, for $L=5$.
    The bold and italic numbers stand for the maximum and minimum values, respectively, for each combination of a range and a repetition $n$, which is a $3\times 3$ matrix.  \label{tab:range_CI_L5}}
  \begin{tabular}{cc|ccc|ccc|ccc}\toprule
    \multicolumn{2}{c|}{$L=5$}                      & \multicolumn{3}{c}{$n=5$} & \multicolumn{3}{|c|}{$n=10$} & \multicolumn{3}{c}{$n=100$}                                                                                                                                      \\
                                                    & $\lambda \backslash p$    & $0.7$                        & $0.9$                       & $0.95$           & $0.7$            & $0.9$            & $0.95$           & $0.7$            & $0.9$            & $0.95$           \\\midrule
    Range of $\hat{p}$                              & $0.05$                    & $0.400$                      & $0.240$                     & $\mathit{0.160}$ & $0.300$          & $0.200$          & $\mathit{0.140}$ & $0.192$          & $0.126$          & $\mathit{0.088}$ \\
    $= \hat{p}_u - \hat{p}_l$                       & $0.1$                     & $0.400$                      & $0.280$                     & $\mathit{0.160}$ & $0.340$          & $0.240$          & $0.160$          & $0.256$          & $0.168$          & $0.122$          \\
                                                    & $0.5$                     & $\mathbf{0.600}$             & $0.360$                     & $0.240$          & $\mathbf{0.580}$ & $0.360$          & $0.240$          & $\mathbf{0.568}$ & $0.346$          & $0.236$          \\\midrule
    Range of $\hat{\sigma}^2_r$                     & $0.05$                    & $0.180$                      & $0.200$                     & $0.140$          & $0.131$          & $0.144$          & $0.111$          & $0.076$          & $0.093$          & $\mathit{0.073}$ \\
    $= \hat{\sigma}^2_{r,u} - \hat{\sigma}^2_{r,l}$ & $0.1$                     & $0.200$                      & $0.180$                     & $0.140$          & $0.147$          & $0.164$          & $0.114$          & $0.099$          & $0.113$          & $0.091$          \\
                                                    & $0.5$                     & $\mathbf{0.220}$             & $0.140$                     & $\mathit{0.100}$ & $\mathbf{0.182}$ & $0.136$          & $\mathit{0.100}$ & $\mathbf{0.160}$ & $0.126$          & $0.093$          \\\midrule
    Range of $\hat{\sigma}^2_L$                     & $0.05$                    & $0.148$                      & $0.072$                     & $\mathit{0.036}$ & $0.083$          & $0.043$          & $\mathit{0.027}$ & $0.033$          & $0.017$          & $\mathit{0.011}$ \\
    $= \hat{\sigma}^2_{L,u} - \hat{\sigma}^2_{L,l}$ & $0.1$                     & $0.172$                      & $0.100$                     & $0.072$          & $0.109$          & $0.060$          & $0.042$          & $0.059$          & $0.037$          & $0.025$          \\
                                                    & $0.5$                     & $\mathbf{0.264}$             & $0.204$                     & $0.200$          & $\mathbf{0.245}$ & $0.201$          & $0.190$          & $\mathbf{0.223}$ & $0.188$          & $0.169$          \\\midrule
    Range of $\hat{\sigma}^2_R$                     & $0.05$                    & $0.160$                      & $0.196$                     & $\mathit{0.148}$ & $0.120$          & $0.155$          & $\mathit{0.123}$ & $0.078$          & $0.101$          & $\mathit{0.080}$ \\
    $= \hat{\sigma}^2_{R,u} - \hat{\sigma}^2_{R,l}$ & $0.1$                     & $0.164$                      & $0.208$                     & $0.148$          & $0.141$          & $0.187$          & $0.139$          & $0.105$          & $0.135$          & $0.110$          \\
                                                    & $0.5$                     & $0.260$                      & $\mathbf{0.268}$            & $0.220$          & $0.255$          & $\mathbf{0.265}$ & $0.213$          & $0.242$          & $\mathbf{0.257}$ & $0.206$          \\\bottomrule
  \end{tabular}
  
  \bigskip
  
  \caption{Ranges covering $95\%$ simulated experiment results of $p$, $\sigma_r^2$, $\sigma_L^2$, and $\sigma_R^2$, for $L=10$.
    The bold and italic numbers stand for the maximum and minimum values, respectively, for each combination of a range and a repetition $n$, which is a $3\times 3$ matrix. \label{tab:range_CI_L10}}
  \centering
  \begin{tabular}{cc|ccc|ccc|ccc}\toprule
    \multicolumn{2}{c|}{$L=10$}                     & \multicolumn{3}{c}{$n=5$} & \multicolumn{3}{|c|}{$n=10$} & \multicolumn{3}{c}{$n=100$}                                                                                                                                      \\
                                                    & $\lambda$ | $p$           & $0.7$                        & $0.9$                       & $0.95$           & $0.7$            & $0.9$            & $0.95$           & $0.7$            & $0.9$            & $0.95$           \\\midrule
    Range of $\hat{p}$                              & $0.05$                    & $0.280$                      & $0.180$                     & $\mathit{0.120}$ & $0.210$          & $0.140$          & $\mathit{0.100}$ & $0.137$          & $0.089$          & $\mathit{0.064}$ \\
    $= \hat{p}_u - \hat{p}_l$                       & $0.1$                     & $0.300$                      & $0.200$                     & $0.140$          & $0.250$          & $0.160$          & $0.120$          & $0.185$          & $0.119$          & $0.086$          \\
                                                    & $0.5$                     & $\mathbf{0.440}$             & $0.280$                     & $0.180$          & $\mathbf{0.420}$ & $0.260$          & $0.180$          & $\mathbf{0.404}$ & $0.252$          & $0.174$          \\\midrule
    Range of $\hat{\sigma}^2_r$                     & $0.05$                    & $0.130$                      & $0.140$                     & $0.110$          & $0.096$          & $0.102$          & $0.082$          & $0.054$          & $0.066$          & $\mathit{0.053}$ \\
    $= \hat{\sigma}^2_{r,u} - \hat{\sigma}^2_{r,l}$ & $0.1$                     & $0.150$                      & $0.140$                     & $0.100$          & $0.103$          & $0.111$          & $0.091$          & $0.070$          & $0.080$          & $0.066$          \\
                                                    & $0.5$                     & $\mathbf{0.160}$             & $0.110$                     & $\mathit{0.080}$ & $\mathbf{0.133}$ & $0.106$          & $\mathit{0.073}$ & $\mathbf{0.115}$ & $0.095$          & $0.069$          \\\midrule
    Range of $\hat{\sigma}^2_L$                     & $0.05$                    & $0.101$                      & $0.050$                     & $\mathit{0.033}$ & $0.057$          & $0.029$          & $\mathit{0.017}$ & $0.023$          & $0.012$          & $\mathit{0.008}$ \\
    $= \hat{\sigma}^2_{L,u} - \hat{\sigma}^2_{L,l}$ & $0.1$                     & $0.115$                      & $0.064$                     & $0.040$          & $0.072$          & $0.042$          & $0.030$          & $0.040$          & $0.025$          & $0.018$          \\
                                                    & $0.5$                     & $\mathbf{0.197}$             & $0.159$                     & $0.118$          & $\mathbf{0.174}$ & $0.147$          & $0.109$          & $\mathbf{0.158}$ & $0.139$          & $0.103$          \\\midrule
    Range of $\hat{\sigma}^2_R$                     & $0.05$                    & $0.115$                      & $0.144$                     & $\mathit{0.110}$ & $\mathit{0.084}$ & $0.111$          & $0.090$          & $\mathit{0.055}$ & $0.072$          & $0.058$          \\
    $= \hat{\sigma}^2_{R,u} - \hat{\sigma}^2_{R,l}$ & $0.1$                     & $0.119$                      & $0.155$                     & $0.123$          & $0.099$          & $0.128$          & $0.108$          & $0.074$          & $0.095$          & $0.077$          \\
                                                    & $0.5$                     & $0.176$                      & $\mathbf{0.212}$            & $0.162$          & $0.171$          & $\mathbf{0.207}$ & $0.158$          & $0.164$          & $\mathbf{0.198}$ & $0.154$          \\\bottomrule
  \end{tabular}
\end{sidewaystable}

Let $L$ be fixed. 
The respective ranges of $p$ and $\sigma_L^2$ attain the minimum and maximum values in each block when $(p,\lambda) = (0.95,0.05)$ and $(0.7,0.5)$, respectively. 
If both $\lambda$ and $n$ are fixed, then these ranges have the minimum and maximum values when $p=0.95$ and $p=0.7$, respectively; whereas, if both $p$ and $n$ are fixed, the ranges have the minimum and maximum values when $\lambda=0.05$ and $\lambda=0.5$, respectively. 
The reasons for the former are that the variance of $\textit{Beta}(a,b)$ becomes larger when $\lambda$ is larger, and the possibility of obtaining diverse $p_i$s values is increased; the reasons for the latter are that the upper bound of $p$ (i.e., $1$) is close to the mean of its estimates, and the range above the mean is necessarily narrower. 
These are both good properties in practice because a large $p$ and small $\lambda$ mean that the measurement method is more sensitive and has less variation, respectively.

The ranges of $\sigma_r^2$ and $\sigma_R^2$, are not as easily summarized. 
If $n$ and $p$ are fixed, the $\lambda$ values representing their minimum and maximum ranges depend on both $n$ and $p$; and in the same way, if $n$ and $\lambda$ are fixed, the $p$ values representing their minimum and maximum ranges depend on both $n$ and $\lambda$. 
These results reflect the combined effect of $p$ being close to $1$, $\sigma_r^2$ being close to $0$, and $n$ being small.

To summarize, we can see that the proposed unbiased estimators have good properties: the higher the accuracy of a given measurement method, the more accurate the estimate of its POD and between-laboratory variance.

Finally, we conclude this subsection with discussion of unrealistic estimates. 
From the definitions of $\sigma_r^2$, $\sigma_L^2$, and $\sigma_R^2$, it is easy to obtain that $0 \leq \sigma_r^2, \sigma_L^2, \sigma_R^2 \leq 1/4$; however, these estimates can deviate from the restrictions.
Indeed, three estimates sometimes exceed $1/4$, and the estimate $\hat{\sigma}_L^2$ is sometimes negative  (cf.\ Tables~\ref{tab:estres_L5} and \ref{tab:estres_L10}). 
Tables~\ref{tab:nums_unre_estim_L5} and \ref{tab:nums_unre_estim_L10} summarize the number of times that such unrealistic estimates were obtained for $L=5$ and $10$, respectively. 
From the tables, we see that if $L$ and $n$ are larger, there is less probability of obtaining such estimates. 
Indeed, comparing the cases of $L=5$ and $10$, where all the parameters except $L$ are identical, we see that the numbers for $L=10$ are always smaller than for $L=5$; whereas, if we compare the cases $n=5$, $10$, and $100$, where all the parameters except $n$ are identical, we see that the numbers for $n=5$ and $n=100$ are largest and smallest, respectively. 
In particular, if $(p,\lambda )=(0.95,0.05)$, then the numbers when the three precision variances exceed $1/4$ are $0$.

\begin{sidewaystable}
  %  \footnotesize
  \caption{The numbers of unrealistic estimates for $L=5$.
    \label{tab:nums_unre_estim_L5}}
  \centering
  \begin{tabular}{cc|ccc|ccc|ccc}\toprule
                                   & $n$           & $5$    & $5$    & $5$    & $10$   & $10$   & $10$   & $100$  & $100$ & $100$  \\
                                   & $p | \lambda$ & $0.7$  & $0.9$  & $0.95$ & $0.7$  & $0.9$  & $0.95$ & $0.7$  & $0.9$ & $0.95$ \\\midrule
    \# in the case                 & $0.05$        & $4058$ & $3200$ & $1903$ & $3802$ & $3895$ & $2869$ & $405$  & $450$ & $537$  \\
    where $\hat{\sigma}^2_L < 0$   & $0.1$         & $3212$ & $2531$ & $1402$ & $2459$ & $2705$ & $1868$ & $113$  & $152$ & $320$  \\
                                   & $0.5$         & $446$  & $346$  & $150$  & $250$  & $290$  & $136$  & $9$    & $107$ & $84$   \\\midrule
    \# in the case                 & $0.05$        & $1838$ & $4$    & $0$    & $722$  & $1$    & $0$    & $29$   & $0$   & $0$    \\
    where $\hat{\sigma}^2_R > 1/4$ & $0.1$         & $2242$ & $24$   & $1$    & $1301$ & $2$    & $1$    & $398$  & $0$   & $0$    \\
                                   & $0.5$         & $3984$ & $503$  & $124$  & $3740$ & $387$  & $95$   & $3602$ & $323$ & $73$   \\\midrule
    \# in the case                 & $0.05$        & $0$    & $0$    & $0$    & $0$    & $0$    & $0$    & $0$    & $0$   & $0$    \\
    where $\hat{\sigma}^2_L > 1/4$ & $0.1$         & $4$    & $0$    & $0$    & $0$    & $0$    & $0$    & $0$    & $0$   & $0$    \\
                                   & $0.5$         & $295$  & $45$   & $5$    & $193$  & $30$   & $12$   & $84$   & $13$  & $3$    \\\midrule
    \# in the case                 & $0.05$        & $1697$ & $12$   & $0$    & $460$  & $0$    & $0$    & $1$    & $0$   & $0$    \\
    where $\hat{\sigma}^2_r > 1/4$ & $0.1$         & $1262$ & $8$    & $0$    & $331$  & $0$    & $0$    & $0$    & $0$   & $0$    \\
                                   & $0.5$         & $42$   & $0$    & $0$    & $8$    & $0$    & $0$    & $0$    & $0$   & $0$    \\\bottomrule
  \end{tabular}
  
  \bigskip
  
  \caption{The numbers of unrealistic estimates for $L=10$.
    \label{tab:nums_unre_estim_L10}}
  \centering
  \begin{tabular}{cc|ccc|ccc|ccc}\toprule
                                   & $n$           & $5$    & $5$    & $5$    & $10$   & $10$   & $10$   & $100$  & $100$ & $100$  \\
                                   & $p | \lambda$ & $0.7$  & $0.9$  & $0.95$ & $0.7$  & $0.9$  & $0.95$ & $0.7$  & $0.9$ & $0.95$ \\\midrule
    \# in the case                 & $0.05$        & $1838$ & $4$    & $0$    & $722$  & $1$    & $0$    & $29$   & $0$   & $0$    \\
    where $\hat{\sigma}^2_L < 0$   & $0.1$         & $3212$ & $2531$ & $1402$ & $2459$ & $2705$ & $1868$ & $113$  & $152$ & $320$  \\
                                   & $0.5$         & $446$  & $346$  & $150$  & $250$  & $290$  & $136$  & $9$    & $107$ & $84$   \\\midrule
    \# in the case                 & $0.05$        & $1838$ & $4$    & $0$    & $722$  & $1$    & $0$    & $29$   & $0$   & $0$    \\
    where $\hat{\sigma}^2_R > 1/4$ & $0.1$         & $2242$ & $24$   & $1$    & $1301$ & $2$    & $1$    & $398$  & $0$   & $0$    \\
                                   & $0.5$         & $3984$ & $503$  & $124$  & $3740$ & $387$  & $95$   & $3602$ & $323$ & $73$   \\\midrule
    \# in the case                 & $0.05$        & $0$    & $0$    & $0$    & $0$    & $0$    & $0$    & $0$    & $0$   & $0$    \\
    where $\hat{\sigma}^2_L > 1/4$ & $0.1$         & $4$    & $0$    & $0$    & $0$    & $0$    & $0$    & $0$    & $0$   & $0$    \\
                                   & $0.5$         & $295$  & $45$   & $5$    & $193$  & $30$   & $12$   & $84$   & $13$  & $3$    \\\midrule
    \# in the case                 & $0.05$        & $1697$ & $12$   & $0$    & $460$  & $0$    & $0$    & $1$    & $0$   & $0$    \\
    where $\hat{\sigma}^2_r > 1/4$ & $0.1$         & $1262$ & $8$    & $0$    & $331$  & $0$    & $0$    & $0$    & $0$   & $0$    \\
                                   & $0.5$         & $42$   & $0$    & $0$    & $8$    & $0$    & $0$    & $0$    & $0$   & $0$    \\\bottomrule
  \end{tabular}
\end{sidewaystable}

Regarding negative estimates, if $\lambda$ is less with the same $p$, there are fewer negative estimates; whereas, if $p$ is less with the same $\lambda$, there are more such estimates. 
In other words, if $\sigma_L^2$ is small, then the probability of obtaining negative estimates is high.

The ISO 5725 rule states that a negative estimate of the between-laboratory variance should be replaced by $0$; and if we follow this rule, estimates that are more than $1/4$ will be replaced by $1/4$. 
However, even, if an estimate of the reproducibility variance exceeds $1/4$ and is replaced by $1/4$, a new issue arises: how to distribute the $1/4$ between the respective estimates of the repeatability and between-laboratory variances. 
With these replacements, of course, the estimators lose their unbiasedness.

\section{Comparisons of precision measures among existing binary methods}
\label{sec:comparison}
This section compares among three existing methods for assessing the precision of binary measurement methods, Wilrich's method~\citep{wilrich2010DeterminationPrecisionQualitativeMeasurement}, accordance and concordance~\citep{langton2002AnalysingCollaborativeTrialsQualitative}, and ORDANOVA~\citep{gadrich2012ORDANOVAAnalysisOrdinalVariation}, and the proposed method.
The details on the three existing methods are described in Appendix~\ref{sec:three_methods}.
Throughout this section, the numbers of laboratories $L$ and repetitions $n$ in a collaborative study are assumed to be determined, and the proofs are omitted because all the propositions in this section are easily derived.

First, the following are relationships between Wilrich's and our precision measures.
\begin{proposition}\label{prop:OursWilrich}
  Assume that the expectation of POD is unknown.
  There exist the following relationships between the definition of precision measures and their estimators proposed by \citet{wilrich2010DeterminationPrecisionQualitativeMeasurement} and that of the present study:
  \begin{equation*}
    %     \label{eq:prop_wilrich:1}
    \sigma^2_r = \sigma^2_{r;W}, \
    \sigma^2_L = \frac{L-1}{L}\sigma^2_{L;W}, \text{ and }
    \sigma^2_R = \sigma^2_{r;W} + \frac{L-1}{L}\sigma^2_{L;W},
  \end{equation*}
  and
  \begin{equation}
    \label{eq:prop_wilrich:2}
    \hat{\sigma}^2_r = \hat{\sigma}^2_{r;W}, \ 
    \hat{\sigma}^2_L = \hat{\sigma}^2_{L;W}, \text{ and }
    \hat{\sigma}^2_R = \hat{\sigma}^2_{R;W},
  \end{equation}
  where the subscript $W$ means it is proposed by \citet{wilrich2010DeterminationPrecisionQualitativeMeasurement}.
\end{proposition}

Since our model assumes a beta distribution as the dispersion followed by $p_i$, the true expectation of $p_i$, $p$, can be supposed; whereas, $p$ cannot be supposed in Wilrich's model, which employs the unbiased estimator $\hat{p}=(1/L) \sum_{i=1}^{L}p_i$, instead of $p$.
Hence, we can use the beta-binomial model in this study when the POD of a measurement method is known in advance; but Wilrich's model cannot be so utilized.
If the expectation of POD is unknown, the estimators of the three precision variances are identical in the two models.
Also, from~\eqref{eq:prop_wilrich:2}, unrealistic estimates can be obtained in both models.

Then, from Proposition~\ref{prop:OursWilrich} and Proposition \ref{prop:LangtonWilrich} in Appendix~\ref{sec:three_methods}, we have relationships between accordance and concordance, proposed by \citet{langton2002AnalysingCollaborativeTrialsQualitative}, and ourselves.
\begin{proposition}
  \label{prop:OursLangton}
  Assume that the expectation of POD is unknown.
  The following relationships exist between the estimators of the precision measures proposed by \citet{langton2002AnalysingCollaborativeTrialsQualitative} and those of the present study:
  \begin{equation*}
    \hat{\sigma}^2_r = \frac{1- \hat{A}}{2}, \quad
    \hat{\sigma}^2_L = \frac{ \hat{A} - \hat{C} }{2}, \quad \text{and} \quad
    \hat{\sigma}^2_R = \frac{1- \hat{C}}{2},
  \end{equation*}
  where $\hat{A}$ and $\hat{C}$ are accordance and concordance, respectively.
\end{proposition}

Since $\hat{\sigma}_r^2$ and $\hat{\sigma}_R^2$ express only accordance and concordance, respectively, we can confirm that accordance and concordance correspond to the repeatability and reproducibility variances, respectively; but if accordance and concordance are close to $1$, then the precision is greater, which is opposite to the case of the precision variances.

Finally, the following are relationships between ORDANOVA's and our precision measures.
\begin{proposition}\label{prop:OursORDANOVA}
  Assume that the expectation of POD is unknown.
  The following relationships exist between the definition of precision measures and their estimators proposed by \citet{gadrich2012ORDANOVAAnalysisOrdinalVariation} and that of the present study:
  \begin{equation*}
    \sigma^2_r = \frac{1}{4}\sigma^2_{r;O}, \quad
    \sigma^2_L = \frac{1}{4}\sigma^2_{L;O}, \quad \text{and} \quad
    \sigma^2_R = \frac{1}{4}\sigma^2_{R:O}
  \end{equation*}
  and
  \begin{align*}
    \hat{\sigma}^2_r & = \frac{n}{4(n-1)}\hat{\sigma}^2_{r;O},                                                     \\
    \hat{\sigma}^2_L & = \frac{L}{4(L-1)}\hat{\sigma}^2_{L;O} - \frac{1}{4(n-1)}\hat{\sigma}^2_{r;O}, \text{ and } \\
    \hat{\sigma}^2_R & = \frac{L}{4(L-1)}\hat{\sigma}^2_{R;O} - \frac{1}{4(L-1)}\hat{\sigma}^2_{r;O},
  \end{align*}
  where the subscript $O$ stands for ORDANOVA and it is proposed by \citet{gadrich2012ORDANOVAAnalysisOrdinalVariation}.
\end{proposition}
The three precision variances in our model are each exactly $1/4$ of ORDANOVA's.
Regarding the estimators, as ORDANOVA expresses the estimators of the three precision variances using only unbiased estimators of the POD of each laboratory $p_i$, and the overall POD $p$, the estimates always take realistic values but do not have unbiasedness.

\section{Laboratory effects: Test methods for homogeneity of proportions}
\label{sec:test-meth-homog}

\subsection{Comparison of four test methods}

Throughout this section, we consider the null hypothesis
\begin{equation}
  \label{eq:H0}
  H_0: \sigma_L = 0 %p_1 = p_2 = \cdots  = p_L.
\end{equation}
%and let $\alpha \in (0,1)$.
A major objective of collaborative studies is to determine whether laboratory effects exist; that is, whether $H_0$ is rejected. 
In order to propose suitable test methods, this section compares four different test methods for homogeneity of proportions: the standard chi-squared, \citet{potthoff1966TestingHomogeneityBinomialMultinomial}, \citet{nass1959X2TestSmallExpectations}, and \citet{xu2011StatisticalIssuesMetaAnalysis} test methods; which are described in detail in Appendix~\ref{sec:test}.
The reasons for these choices are as follows. 
The standard chi-squared method is well-known and well-used. 
Potthoff and Whittinghill's test method would be suitable for beta-binomial models because it assumes, as the alternative hypothesis, that the distribution $p_1,\ldots,p_L$ is independent and identically follows a beta distribution $\textit{Beta}(cp,c (1-p))$ for some $c>0$, and that each $x_i \ (1\leq i\leq L)$ follows a binomial distribution $\textit{Bi}(n,p_i)$.
Finally, a previous study~\citep{klein2013ComparisonTestsHomogeneityBinomial} reported that, out of nine test methods, Nass's and Xu's test methods tended to perform adequately in their various simulation environments, and concluded that these seemed appropriate for testing the homogeneity of binomial proportions.

To evaluate and compare the four statistical tests, we derived statistical powers using simulated collaborative studies, as in Subsection~\ref{sec:sim_study}.
The simulated studies assumed that the same beta-binomial models as in Subsection~\ref{sec:sim_study} held as the alternative hypotheses. 
For each case, the number of laboratories $L$ was set at $5$ and $10$, and the number of repetitions $n$ was set at $5$, $10$, and $100$. 
Each of these collaborative studies was repeated $10,000$ times.

For each simulated collaborative study, we conducted the four tests (standard chi-squared, Potthoff and Whittinghill, Nass, and Xu), at a significance level of $5\%$, and then calculated the probability of studies detecting laboratory effects for each test method as a statistical power. 
Tables~\ref{tab:test_power_L5} and \ref{tab:test_power_L10} show the power of the four test methods for $L=5$ and $L=10$, respectively. 
In the tables, the bold and italicized numbers respectively indicate the maximum and minimum probabilities for the same $p$, $\lambda$, and $n$ values.

\begin{sidewaystable}
  \caption{Probability of studies detecting the laboratory effect in the four methods, for $10000$ simulated collaborative studies, when $L=5$.
    Std, PW, Nass, and Xu refer to the standard chi-squared, Potthoff and Whittinghill, Nass, and Xu test methods, respectively. 
    In the table, the bold and italic numbers indicate the maximum and minimum probabilities for the same $p$, $\lambda$, and $n$. 
    \label{tab:test_power_L5}}
  \centering
  \begin{tabular}{cc|cccc|cccc|cccc}\toprule
           &                               & \multicolumn{4}{c}{$n=5$} & \multicolumn{4}{|c|}{$n=10$} & \multicolumn{4}{c}{$n=100$}                                                                                                                                                                   \\
    $p$    & $\lambda$ $\backslash$ Method & Std                       & PW                           & Nass                        & Xu               & Std     & PW               & Nass             & Xu               & Std              & PW               & Nass             & Xu               \\\midrule
    %    $0.7$ & - & $0.037$ & $0.027$ & $0.042$ & $0.052$ & $0.046$ & $0.027$ & $0.054$ & $0.061$ & $0.048$ & $0.026$ & $0.048$ & $0.053$\\\hline
    $0.7$  & $0.05$                        & $0.082$                   & $\mathit{0.061}$             & $0.093$                     & $\mathbf{0.105}$ & $0.153$ & $\mathit{0.109}$ & $0.166$          & $\mathbf{0.180}$ & $0.822$          & $\mathit{0.780}$ & $0.823$          & $\mathbf{0.831}$ \\
    $0.7$  & $0.1$                         & $0.134$                   & $\mathit{0.107}$             & $0.148$                     & $\mathbf{0.163}$ & $0.303$ & $\mathit{0.240}$ & $0.320$          & $\mathbf{0.342}$ & $0.942$          & $\mathit{0.924}$ & $0.942$          & $\mathbf{0.945}$ \\
    $0.7$  & $0.5$                         & $0.671$                   & $\mathit{0.618}$             & $\mathbf{0.701}$            & $0.687$          & $0.850$ & $\mathit{0.816}$ & $\mathbf{0.872}$ & $0.866$          & $\mathbf{0.995}$ & $\mathit{0.992}$ & $\mathbf{0.995}$ & $\mathbf{0.995}$ \\\midrule
    %    $0.9$ & - & $0.022$ & $0.009$ & $0.066$ & $0.023$ & $0.030$ & $0.013$ & $0.069$ & $0.056$ & $0.051$ & $0.024$ & $0.051$ & $0.057$\\\hline
    $0.9$  & $0.05$                        & $0.052$                   & $\mathit{0.026}$             & $\mathbf{0.113}$            & $0.054$          & $0.108$ & $\mathit{0.067}$ & $\mathbf{0.183}$ & $0.159$          & $0.798$          & $\mathit{0.737}$ & $0.800$          & $\mathbf{0.808}$ \\
    $0.9$  & $0.1$                         & $0.087$                   & $\mathit{0.046}$             & $\mathbf{0.158}$            & $0.091$          & $0.193$ & $\mathit{0.136}$ & $\mathbf{0.289}$ & $0.253$          & $0.916$          & $\mathit{0.884}$ & $0.917$          & $\mathbf{0.921}$ \\
    $0.9$  & $0.5$                         & $0.337$                   & $\mathit{0.301}$             & $\mathbf{0.429}$            & $0.339$          & $0.494$ & $\mathit{0.443}$ & $\mathbf{0.577}$ & $0.512$          & $0.834$          & $\mathit{0.810}$ & $\mathbf{0.855}$ & $0.834$          \\\midrule
    %   $0.95$ & - & $0.005$ & $0.002$ & $0.039$ & $0.005$ & $0.010$ & $0.003$ & $0.073$ & $0.025$ & $0.032$ & $0.004$ & $0.033$ & $0.036$\\\hline
    $0.95$ & $0.05$                        & $0.020$                   & $\mathit{0.010}$             & $\mathbf{0.084}$            & $0.020$          & $0.061$ & $\mathit{0.034}$ & $\mathbf{0.171}$ & $0.096$          & $0.749$          & $\mathit{0.667}$ & $0.752$          & $\mathbf{0.755}$ \\
    $0.95$ & $0.1$                         & $0.047$                   & $\mathit{0.027}$             & $\mathbf{0.125}$            & $0.047$          & $0.123$ & $\mathit{0.077}$ & $\mathbf{0.257}$ & $0.163$          & $0.834$          & $\mathit{0.775}$ & $\mathbf{0.843}$ & $0.836$          \\
    $0.95$ & $0.5$                         & $0.190$                   & $\mathit{0.174}$             & $\mathbf{0.262}$            & $0.190$          & $0.303$ & $\mathit{0.255}$ & $\mathbf{0.376}$ & $0.312$          & $0.595$          & $\mathit{0.561}$ & $\mathbf{0.632}$ & $0.595$          \\\bottomrule
  \end{tabular}
  
  \bigskip
  
  \caption{
    Probability of studies detecting the laboratory effect in the four methods, for $10000$ simulated collaborative studies, when $L=10$. 
    Std, PW, Nass, and Xu refer to the standard chi-squared, Potthoff and Whittinghill, Nass, and Xu test methods, respectively. 
    In the table, the bold and italic numbers indicate the maximum and minimum probabilities for the same $p$, $\lambda$, and $n$.
    \label{tab:test_power_L10}}
  \centering
  \begin{tabular}{cc|cccc|cccc|cccc}\toprule
           &                               & \multicolumn{4}{c}{$n=5$} & \multicolumn{4}{|c|}{$n=10$} & \multicolumn{4}{c}{$n=100$}                                                                                                                                                                   \\
    $p$    & $\lambda$ $\backslash$ Method & Std                       & PW                           & Nass                        & Xu               & Std     & PW               & Nass             & Xu               & Std              & PW               & Nass             & Xu               \\\midrule
    $0.7$  & $0.05$                        & $0.108$                   & $\mathit{0.089}$             & $0.127$                     & $\mathbf{0.147}$ & $0.240$ & $\mathit{0.195}$ & $0.251$          & $\mathbf{0.278}$ & $0.976$          & $\mathit{0.969}$ & $0.976$          & $\mathbf{0.979}$ \\
    $0.7$  & $0.1$                         & $0.200$                   & $\mathit{0.173}$             & $0.225$                     & $\mathbf{0.254}$ & $0.488$ & $\mathit{0.432}$ & $0.500$          & $\mathbf{0.533}$ & $\mathbf{0.999}$ & $\mathbf{0.999}$ & $\mathbf{0.999}$ & $\mathbf{0.999}$ \\
    $0.7$  & $0.1$                         & $0.908$                   & $\mathit{0.882}$             & $0.918$                     & $\mathbf{0.926}$ & $0.986$ & $\mathit{0.980}$ & $\mathbf{0.987}$ & $\mathbf{0.987}$ & $\mathbf{1.000}$ & $\mathbf{1.000}$ & $\mathbf{1.000}$ & $\mathbf{1.000}$ \\\midrule
    $0.9$  & $0.05$                        & $0.094$                   & $\mathit{0.047}$             & $0.108$                     & $\mathbf{0.127}$ & $0.206$ & $\mathit{0.137}$ & $0.218$          & $\mathbf{0.219}$ & $0.966$          & $\mathit{0.956}$ & $0.967$          & $\mathbf{0.971}$ \\
    $0.9$  & $0.1$                         & $0.180$                   & $\mathit{0.111}$             & $0.202$                     & $\mathbf{0.224}$ & $0.396$ & $\mathit{0.302}$ & $0.410$          & $\mathbf{0.411}$ & $0.996$          & $\mathit{0.994}$ & $0.996$          & $\mathbf{0.997}$ \\
    $0.9$  & $0.5$                         & $0.642$                   & $\mathit{0.554}$             & $0.647$                     & $\mathbf{0.653}$ & $0.819$ & $\mathit{0.773}$ & $\mathbf{0.826}$ & $\mathbf{0.826}$ & $0.979$          & $\mathit{0.971}$ & $\mathbf{0.980}$ & $0.979$          \\\midrule
    $0.95$ & $0.05$                        & $0.083$                   & $\mathit{0.028}$             & $0.086$                     & $\mathbf{0.092}$ & $0.187$ & $\mathit{0.105}$ & $\mathbf{0.198}$ & $\mathbf{0.198}$ & $0.949$          & $\mathit{0.930}$ & $0.950$          & $\mathbf{0.954}$ \\
    $0.95$ & $0.1$                         & $0.145$                   & $\mathit{0.066}$             & $0.150$                     & $\mathbf{0.158}$ & $0.334$ & $\mathit{0.221}$ & $\mathbf{0.346}$ & $\mathbf{0.346}$ & $0.981$          & $\mathit{0.970}$ & $0.982$          & $\mathbf{0.983}$ \\
    $0.95$ & $0.5$                         & $0.434$                   & $\mathit{0.327}$             & $0.435$                     & $\mathbf{0.437}$ & $0.594$ & $\mathit{0.524}$ & $\mathbf{0.602}$ & $\mathbf{0.602}$ & $0.867$          & $\mathit{0.836}$ & $\mathbf{0.869}$ & $0.867$          \\\bottomrule
  \end{tabular}
\end{sidewaystable}

The standard chi-squared test method shows intermediate probabilities among the four test methods.
For the case where $n=100$, the probabilities are close to the highest among the four methods.
Therefore, this method can be appropriate for cases with a large number of repetitions; and a well-known property, that in large sample groups the accuracy of the approximation to the standard chi-squared distribution becomes high, is confirmed in the beta-binomial models.

The Potthoff and Whittinghill test method records the minimum probabilities in all the environments, even though it assumes a beta-binomial distribution as the alternative hypothesis.
Hence, this method is not recommended for detecting laboratory effects in the beta-binomial models.

In all the environments, either the Nass or Xu test method records the maximum probabilities. 
In the case of $L=5$, if $p=0.90, 0.95$ and $n=5,10$, then the Nass test records the maximum probabilities; and if p=$0.70$, then both methods show the maximum probabilities when $\lambda=0.5$ and $\lambda=0.05,0.1$, respectively.

\subsection{Further comparison of Nass and Xu test methods}

Based on the observations in the previous subsection, the present study conducted further simulated studies to consider which of the Nass and Xu test methods can be recommended. 
The following cases were treated:
$(a, b) = (0.7, 0.3)$, $(6.3, 2.7)$, $(13.3, 5.7)$, $(69.3, 29.7)$, $(139.3, 59.7)$, $(0.8, 0.2)$, $(7.2, 1.8)$, $(15.2, 3.8)$, $(79.2, 19.8)$, $(159.2, 39.8)$, $(0.9, 0.1)$, $(8.1, 0.9)$, $(17.1, 1.9)$, $(89.1, 9.9)$, $(179.1, 19.9)$, $(0.95, 0.05)$, $(8.55, 0.45)$, $(18.05, 0.95)$, $(94.05, 4.95)$, $(189.05, 9.95)$ The expectations of $Beta(a, b)$ in the first, second, third, and last four cases are $0.7$, $0.8$, $0.9$, and $0.95$, respectively, and the over-dispersion parameters of $BBi(n, a, b)$ of each group are $0.5$, $0.1$, $0.05$,  $0.01$, and  $0.005$, respectively. 
For each case, the number of laboratories $L$ was set from $3$ to $15$, and the number of measured values in each laboratory $n$ was set at $3, 4, 5, 6, 7, 8, 9, 10, 15, 20, 25, 30, 35, 40, 45, 50, 60, 70, 80,
  90$, and $100$. 
Each of these collaborative studies was repeated $1,000$ times.

The Nass test method is theoretically suitable when the observed values are sparse, i.e., when $nq:=n \min\{p, (1-p)\}$ is small.
Conversely, the Xu test method is appropriate when $nL$ is large.
Thus, the present study summarizes the relationships between the test method that rejected more often and the value of $nLq$. 
Note that when one test method was rejected 10 or more times, that is, $1\%$ or more of the repeated times, than the other, it was deemed that the method rejected more often.

\begin{figure}
  \begin{center}
    \includegraphics[scale=0.5]{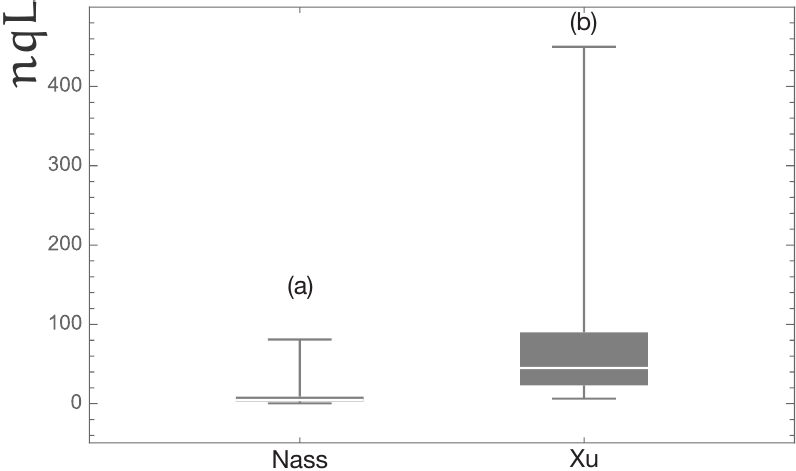}
  \end{center}
  \caption{
    Box-and-whisker plot showing (a) the values of $nqL$ for which the Nass test method rejected 10 or more times than the Xu test method , and (b) the values of $nqL$ for which the Xu test method rejected 10 or more times than the the Nass test method.
    \label{fig:boxchart}}
\end{figure}

\begin{figure}
  \begin{center}
    \includegraphics[scale=0.6]{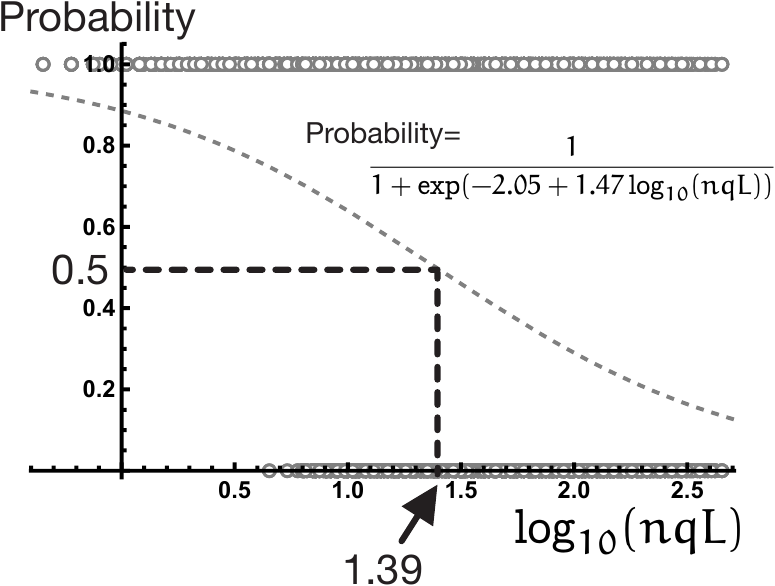}
  \end{center}
  \caption{
    Logistic regression model with $\log_{10} (qnL)$  as the explanatory variable to predict the probability that the Nass test method has a better statistical power. 
    \label{fig:logistic}}
\end{figure}

Figure~\ref{fig:boxchart} shows the results using a Box-and-whisker plot.
It confirms the trend that the Nass test method exhibits better statistical power when $nqL$ is small, while the Xu test method is more effective  when $nqL$ is large.
To provide a practical threshold for determining the recommended test method, the present study applyed a logistic regression model with $\log_{10} (quL)$ as the explanatory variable.
The response value $1$ and $0$ indicate that the Nass and Xu test methods rejected 10 or more times than the other, respectively, and for the cases with differences less than 10 times, both $0$ and $1$ were assigned as response values.
The logistic regression model was predicted as $\text{Probability} = (1+ \exp(-2.05 + 1.47 \log_{10} (qnL))^{-1}$, and the $50\%$ point of $\log_{10} (quL)$ was about $1.39$ (Figure~\ref{fig:logistic}).
Therefore, we recommend using the Nass method when $nqL < 25 (\simeq 10^{1.39})$, and the Xu method in other cases.

% However, in each setting, the difference between the probabilities in the Nass and Xu tests is not very large, ranging from $0.01$ to $0.02$. 
% Regarding the case where $L=10$, if $n=5,10$, then the Xu test records the maximum probabilities. 
% Based on these observations, the present study recommends using the Nass method when both $n$ and $L$ are very small (less than $10$), and the Xu method in other cases. 
% Finally, we note that, in the case of $n=100$, it is difficult to determine which of the two test types is more dominant; however, as it is rare to have such large repetitions in realistic collaborative studies, the above conclusion was drawn.

\section{Real example: Listeria monocytogenes}
\label{sec:ex}
%\label{sec:listeria}

This subsection analyzes the results of a collaborative study on \textit{Listeria monocytogenes}, which was presented in ISO 16140~\citep{iso2016ISO161402016Microbiology} and analyzed by \citet{wilrich2010DeterminationPrecisionQualitativeMeasurement}.
The study involved ten laboratories, and each laboratory repeated the measurements five times; that is, $(L,n)=(10,5)$.
The results are shown in Table~\ref{tbl:ex_Listeria}. 
In the table, the first, second, and last columns show the laboratory number, the detected results for \textit{L.\ monocytogenes}, and the number of detections, respectively. 
Here, $y_{ij}=1$ and $0$ indicate that \textit{L.\ monocytogenes} was and was not detected, respectively.

\begin{table}[htbp]
  \caption{Measured values of a collaborative study on \textit{L. monocytogenes}.
    In the second column of the table, $y_{ij}=1$ and $0$ indicate that the bacterium was and was not detected, respectively.
    \label{tbl:ex_Listeria}}
  \centering
  \begin{tabular}{ccc}\toprule
    Laboratory $i$ & Measured values $y_{ij}$ & $\displaystyle{x_i = \sum_{j = 1}^n y_{ij}}$ \\\midrule
    1              & $1,1,1,1,1$              & 5                                            \\
    2              & $1,1,1,1,1$              & 5                                            \\
    3              & $1,1,1,1,1$              & 5                                            \\
    4              & $1,1,1,1,1$              & 5                                            \\
    5              & $0,0,1,1,1$              & 3                                            \\
    6              & $1,1,1,1,1$              & 5                                            \\
    7              & $0,0,1,1,1$              & 3                                            \\
    8              & $1,1,1,1,1$              & 5                                            \\
    9              & $1,1,1,1,1$              & 5                                            \\
    10             & $1,1,1,1,1$              & 5                                            \\\bottomrule
  \end{tabular}
\end{table}

From Proposition~\ref{prop:est}, we obtain the following estimates:
\begin{align*}
  \hat{p}_i      & =
  \begin{cases}
    1.0  & (i = 1,2,3,4,6,8,9,10), \\
    0.60 & (i= 5,7),
  \end{cases}                                                         \\
  \hat\sigma^2_r & = 0.060, \hat\sigma^2_L  = 0.016, \text{ and } \hat\sigma^2_R  = 0.076.
\end{align*}
Since $p$ is the expectation of $p_i$, we have $\hat{p}=0.92$.
To determine whether the laboratory effect exists, the Nass test is conducted because $n\hat{q}L = 10 \times 0.08 \times 5 =4 < 25$, where $\hat{q}:=\min\{\hat{p}, 1- \hat{p}\}$.
Since the POD is unknown in this example, its unbiased estimate $\hat{p}$ is used instead of $p$.
The test statistic and the critical value are respectively $26.2$ and $23.4$; therefore, the laboratory effect is detected.
Note that if the standard test is conducted, then the test statistic and the critical value are $17.4$ and $16.9$, respectively.
Hence, the laboratory effect is detected, which coincides with the result of the Nass test.

\section{Conclusions}
\label{sec:conc}

The present study introduced a new model for evaluating the precision of binary measurement methods. 
In addition, it defined the precision measures of repeatability and reproducibility for the model, and provided their unbiased estimators. 
The key concept was the assumption of beta-binomial data distributions in collaborative studies. 
From simulated studies, a good property of the unbiased estimators was found: the higher the accuracy of a given measurement method, the more accurate the estimates of its POD and between-laboratory variance.

Next, the study derived the formulae for conversion from the precision measures and their estimators in existing models \citep{wilrich2010DeterminationPrecisionQualitativeMeasurement,langton2002AnalysingCollaborativeTrialsQualitative,gadrich2012ORDANOVAAnalysisOrdinalVariation} to those of the present study. 
In particular, the estimators of \citet{wilrich2010DeterminationPrecisionQualitativeMeasurement} and the present study were found to be identical when the expectation of POD was unknown, even though the assumptions of the two models were different. 
Also, the precision measures of ORDANOVA~\citep{gadrich2012ORDANOVAAnalysisOrdinalVariation} were exactly $4$ times those of the present study, but these estimators were different.
The advantage of the ORDANOVA estimators lay in their never taking unrealistic values, but their good statistical properties were not observed.

Further, based on the simulated studies and statistical analysis, we proposed appropriate statistical tests to detect laboratory effects. 
In sum, the proposal was to use the Nass test method when $n q L  < 25 \ (q=\min\{p, 1-p\})$ and the Xu test method in other cases. 
We here should note that all the four statistical test methods are based on comparisons of
the between-laboratory and reproducibility variances. 
Xu's test method is used an unbiased estimate of the between-laboratory variance; on the other hand, the standard chi-squared test and its extended test methods (Potthoff and Whittinghill's test and Nass's test method) are not used unbiased estimates. 
The effects of the unbiasedness on statistical test methods were discussed in \citet{gadrich2013Comparisonbiasedunbiased}.

We applied the proposed test procedure to a real-world example from food safety assessment.
However, though studies such as those of Wilrich, and Gadrich and Bashkansky, utilized the standard chi-squared test, the present study could not use this as part of its own proposed method, because it focused on collaborative studies with small numbers of participating laboratories and repetitions.
However, such collaborative studies are sometimes conducted, especially in the biological field. 
We also remark that the collaborative study analyzed here was a real-world example that provided important information for discussions in a public institution, the International Organization for Standardization (ISO).

Finally, we note some further challenges. 
First, Proposition~\ref{prop:est} proved that the proposed estimators of repeatability and reproducibility variances were unbiased. 
While this is one of the good properties for estimators, unbiased estimators are generally not unique. 
If the efficiency of the unbiased estimators can be proven, then our proposed estimators are more appropriate; but proofs are not given. 
Also, the proposed estimators can take unrealistic values, as discussed in Subsection~\ref{sec:sim_study}. 
In the case of quantitative data for which a normal distribution can be assumed, in addition to the ANOVA-based estimator, the restricted maximum likelihood (REML) estimator \citep{thompson1962ProblemNegativeEstimatesVariance,harville1977MaximumLikelihoodApproachesVariance} is used. 
Although the present study extended an ANOVA-based approach, it may be possible to consider extending the REML estimator to qualitative data.
In addition, providing expressions for the variances of estimators would be crucial to emphasize the utility of the proposed estimators, but this is a future challenge.

Second, we proposed using the Nass or Xu test method for detecting laboratory effects based on the simulated studies; however, ideally, we should recommend the choice between Nass and Xu methods based on theoretical and mathematical considerations.
This remains a topic for further work.
Also, the powers are not high enough if the numbers of participating laboratories and their repetitions are small, and such situations are common in biological fields. 
Thus, we need to consider whether our proposed methods and results are sufficient for decision-making in such areas. 

Third, the reproducibility variances in the study depended solely on $p_i$; and this implies that, for any measurement method, only POD determines their reproducibility. 
In other words, the proposed model cannot assess the difference between two measurement methods with the same POD but different precision. 
This problem, which is a well-known property of models based on binomial distributions, is not solved by the proposed model.

\part*{Appendices}
%\begin{appendices}

\section{Proofs}
\label{sec:proofs}

\subsection{Proof of Proposition~\ref{prop:variances-theoritical}}
\label{ssec:pr-variances-theoritical}

\begin{proof}
  Since $y_{ij}|p_i$ follows a Bernoulli distribution $\textit{Be} (p_i)$, the repeatability variance  in laboratory $i$ is $\sigma^2_{ri} = p_i (1-p_i)$.
  Therefore, as $p_i$ follows a beta distribution $\textit{Beta} (a, b)$, the repeatability variance $\sigma^2_r = E(\sigma^2_{ri})$ is as follows:
  \begin{align}
    \sigma^2_r 
     & = E( p_i (1 - p_i)) \nonumber                                                                    \\
     & = E(p_i) - \left(E(p_i^2)\right) = E(p_i) - \left(V(p_i) + \left(E(p_i)\right)^2\right)\nonumber \\
     & = \frac{a b}{(a + b)(a + b + 1)}\label{eq:repetability-alphabeta}.
  \end{align}
  
  Next, we consider the meaning of between-laboratory variance, $\sigma^2_L = V(p_i)$.
  Since $p_i$ follows a beta distribution $Beta(a, b)$, we have:
  \begin{equation}
    \sigma^2_L = \frac{a b}{(a + b)^2 (a + b + 1)}.
    \label{eq:betweenlab-alphabeta}
  \end{equation}
  
  Finally, from \eqref{eq:repetability-alphabeta} and \eqref{eq:betweenlab-alphabeta}, and the definition of reproducibility variance, $\sigma^2_R = \sigma^2_r + \sigma^2_L$ , we have:
  \begin{equation*}
    \sigma^2_R = \frac{a b}{(a + b)(a + b + 1)}
    +  \frac{a b}{(a + b)^2 (a + b + 1)}
    = \frac{a b}{(a + b)^2}.
    %\label{eq:reproducibility-alpha-beta}
  \end{equation*}
\end{proof}

\subsection{Proof of Proposition~\ref{prop:variances-compare}}
\label{ssec:pr-variances-compare}

\begin{proof}
  Since $\sigma^2_{\textit{BBi}}$ is the variance of a beta-binomial distribution $\textit{BBi} (n, a, b)$, 
  \begin{equation}
    \sigma^2_{\textit{BBi}} = \frac{n a b ( a + b + n )}{(a + b)^2 (a + b + 1)}
    = n \frac{a b }{(a+b)(a + b + 1)}
    + n^2 \frac{a b}{(a + b)^2 ( a + b + 1 )}.
    \label{eq:BBi-alphabeta}
  \end{equation}
  From \eqref{eq:repetability-alphabeta}, \eqref{eq:betweenlab-alphabeta}, and \eqref{eq:BBi-alphabeta}, we have:
  \begin{equation}
    \sigma^2_{\textit{BBi}} = n \sigma^2_r + n^2 \sigma^2_L.
    \label{eq:relation-variances1}
  \end{equation}
  Further, from the definition of reproducibility variance, we have:
  \begin{equation}
    \sigma^2_R = \sigma^2_r + \sigma^2_L.
    \label{eq:relation-variances2}
  \end{equation}
  
  By solving the simultaneous equations~\eqref{eq:relation-variances1} and \eqref{eq:relation-variances2} on $\sigma^2_L$ and $\sigma^2_R$, we conclude the proposition.
  % Finally, from \eqref{eq:relation-variances2} and $E(p_i) = a / (a + b)$, we conclude the expression~\eqref{eq:variances-sigmaR}.
\end{proof}

\subsection{Proof of Proposition~\ref{prop:est}}\label{sec:proof-est}

\begin{proof}
  %It is well-know that \eqref{eq:unbiased_est_pi} and \eqref{eq:unbiased_est_p} are unbiased estimates.
  It is well known that \eqref{eq:unbiased_est_pi} is an unbiased estimator of POD $p_i$.
  From the definition of $\sigma^2_R$, \eqref{eq:unbiased_est_sigmaR} is an unbiased estimator if \eqref{eq:unbiased_est_sigmar} and \eqref{eq:unbiased_est_sigmaL} are unbiased estimators.
  Therefore, this subsection only proves that \eqref{eq:unbiased_est_sigmar} and \eqref{eq:unbiased_est_sigmaL} are unbiased estimators.
  
  First, we prove that \eqref{eq:unbiased_est_sigmar} is an unbiased estimator.
  From $\hat{p}_i = (1/n) \sum_{i=1}^n y_{ij} = x_i /n$, we have
  \begin{align}
    E\left( \hat{\sigma}_r^2 \right)
     & = E \left(\frac{n}{L(n-1)} \sum_{i=1}^L \hat{p}_i ( 1- \hat{p}_i) \right)\nonumber                              \\
     & =\frac{n}{L(n-1)} \sum_{i=1}^L \left( E(\hat{p}_i) - E(\hat{p}_i^2)\right)\nonumber                             \\
     & = \frac{n}{L(n-1)} \sum_{i=1}^L \left( E(\hat{p}_i) - V(\hat{p}_i)- \left(E(\hat{p}_i)\right)^2\right)\nonumber \\ 
     & = \frac{1}{nL(n-1)} \sum_{i=1}^L \left( nE(x_i) - V(x_i)- \left(E(x_i)\right)^2\right).
    \label{eq:pr_unbiased_est_sigmar1}
  \end{align}
  Since $x_{i}$ follows a beta-binomial distribution $\textit{BBi} (n, a, b)$, we obtain
  \begin{align*}
    \eqref{eq:pr_unbiased_est_sigmar1} & = \frac{1}{nL(n-1)} \sum_{i=1}^L 
    \left( \frac{n^2 a}{a + b}
    - \frac{n a b}{(a + b)^2 (a + b + 1)}
    - \left(\frac{n a}{a + b} \right)^2
    \right)                                                                               \\
                                       & = \frac{a b}{(a + b) (a + b + 1)} = \sigma^2_r.
  \end{align*}
  
  Next, we prove that \eqref{eq:unbiased_est_sigmaL} is an unbiased estimator.
  Since $\hat\sigma^2_r$ and $\hat\sigma^2_{BBi}$ are unbiased estimators, the following holds immediately:
  \begin{align*}
    E(\hat\sigma^2_{L}) & = \frac{E(\hat\sigma^2_{\textit{BBi}}) - n E(\hat\sigma^2_r)}{n^2} \\
                        & = \frac{\sigma^2_{\textit{BBi}} - n \sigma^2_r}{n^2}
    = \sigma^2_{L}.
  \end{align*}
\end{proof}

\section{Existing methods for assessing the precision of binary measurement methods}
\label{sec:three_methods}
\subsection{ISO 5725-based method for binary measurement methods}

This subsection deals with an ISO 5725-based method, originally proposed by~\citet{wilrich2010DeterminationPrecisionQualitativeMeasurement}. 

The basic model to analyze binary measured values is the same as in the ISO 5725 series, rewritten as
\begin{equation*}
  %	\label{eq:wilrich}
  y_{ij} = p + (p_i - p) + e_{ij} \quad \text{with } B_i:= p_i - p,
\end{equation*}
where $y_{ij}$ is the measured value defined as $0$ (negative) or $1$ (positive) for trial $j \in \{ 1,\ldots,n \}$ at laboratory $i \in \{1,\ldots,L\}$; $p_i$ is the POD (i.e., the probability of obtaining a measured value $y_{ij}=1$, for laboratory $i$); and $p$ is its expectation.
Under the assumption that $y_{ij}$ in laboratory $i$ followed a Bernoulli distribution with expectation $p_i$, the within-laboratory variance is defined as $\sigma_{ri;W}^2 := p_i (1-p_i)$.
The repeatability variance is defined as the average of the within-laboratory variances; that is,
\begin{equation*}
  \sigma^2_{r;W}:= \frac{1}{L} \sum_{i=1}^{L} \sigma^2_{ri;W} = \frac{1}{L} \sum_{i=1}^L p_i (1-p_i).  
\end{equation*}
The between-laboratory variance is defined by the variation among the PODs $p_i$; that is,
that is,
\begin{equation*}
  \sigma^2_{L;W} := \frac{1}{L-1}\sum_{i=1}^L \left(p_i - \frac{1}{L} \sum_{i=1}^L p_i\right)^2.
\end{equation*}
Finally, the reproducibility variance is defined as in ISO 5725-2:
\begin{equation*}
  \sigma^2_{R;W} := \frac{1}{L} \sum_{i=1}^{L} p_i (1-p_i) + \frac{1}{L-1}\sum_{i=1}^L \left(p_i - \frac{1}{L} \sum_{i=1}^L p_i \right)^2.
\end{equation*}

To estimate the repeatability, between-laboratory, and reproducibility variances, a one-way ANOVA (a random effects model) is performed and Wilrich proposed the following estimators:
\begin{align*}
  \hat{p}_i            & = \frac{1}{n} \sum_{j=1}^n y_{ij},\quad \hat{p} = \frac{1}{L} \sum_{i=1}^L \hat{p}_i,                                    \\%\label{eq:5725_est_p}\\
  \hat{\sigma}_{r;W}^2 & = \frac{n \sum_{i=1}^L \hat{p}_i (1 - \hat{p_i})}{L(n-1)},                                                               \\%\label{eq:5725_est_sigmar}\\
  \hat{\sigma}_{L;W}^2 & = \frac{\sum_{i=1}^L (\hat{p}_i - \hat{p})^2}{L-1}- \frac{\sum_{i=1}^L \hat{p}_i ( 1- \hat{p}_i) }{L(n-1)}, \text{ and } \\
  %	\label{eq:5725_est_sigmaL}\\
  \hat{\sigma}_{R;W}^2 & = \hat\sigma^2_{r;W} + \hat\sigma^2_{L;W}.
  %	\label{eq:unbiased_est_sigmaR}
\end{align*}

\subsection{Accordance and concordance}

This subsection deals with accordance and concordance, originally proposed by~\citet{langton2002AnalysingCollaborativeTrialsQualitative}.
These measures correspond to repeatability and reproducibility, respectively, in ISO 5725, and are based on the probability that two measured values are identical. 
The definition of accordance is the probability that pairs in each laboratory are identical, while concordance is the probability that pairs between different laboratories are identical. 
The respective estimators of accordance $\hat{A}$ and concordance $\hat{C}$ are
\begin{align*}
  \hat{A} & := \frac{1}{L} \sum_{i=1}^L \hat{A}_i,                          \\
  \hat{C} & := \frac{2 X (X -nL) + nL (nL -1)-\hat{A} nL(n-1)}{n^2L(L-1)},
\end{align*}
where
\begin{equation*}
  \hat{A}_i = \frac{x_i (x_i-1)+(n-x_i)(n-x_i-1)}{n(n-1)} \text{ and } X = \sum_{i=1}^L x_i.
\end{equation*}

These concepts differ greatly between the present study and that of~\citet{wilrich2010DeterminationPrecisionQualitativeMeasurement}; however, we may observe some relationships among the estimators.
The following relationships between Langton et al.'s and Wilrich's precision measures hold (see e.g., \citet{iso2021ISOTR27877Statistical}).
\begin{proposition}
  \label{prop:LangtonWilrich}
  It is the case that
  \begin{equation*}
    \hat{\sigma}^2_{r;W} = \frac{1- \hat{A}}{2}, \quad 
    \hat{\sigma}^2_{L;W} = \frac{\hat{A} - \hat{C}}{2}, \quad \text{and} \quad
    \hat{\sigma}^2_{R;W} = \frac{1- \hat{C}}{2}.
  \end{equation*}
\end{proposition}

\subsection{ORDANOVA for binary measurement methods}
\label{sec:ordanova-binary-case}

ORDANOVA was originally introduced by~\citet{gadrich2012ORDANOVAAnalysisOrdinalVariation} for general ordinal-scale measurement methods; but this subsection deals only with binary cases. 
The basic idea of ORDANOVA is to use an ordinal dispersion measure, proposed by~\citet{blair1996MeasuresVariationOrdinalData,blair2000StatisticsOrdinalVariation}, to express within-laboratory precision. 
For binary data assumed to follow a binomial distribution with parameters $n$ and $q$, the dispersion measure is defined as $\sigma=4q(1-q)$.
Note that $q$ is used here instead of the usual $p$ because $p$ is used for the expectation of $p_i$ in the present study.

Let $y_{ij}$ be the measured value defined as $0$ and $1$ for trial $j \in \{1,\ldots,n\}$ at laboratory $i \in \{1,\ldots,L\}$, and $p_i$ be the POD of laboratory $i$.
Then, under the assumption that each measured value at laboratory $i$ followed a binomial distribution with expectation $p_i$, the within-laboratory variance of laboratory $i$ is defined as $\sigma_{ri;O}^2 := 4p_i (1-p_i)$; and the total variance is defined as $\sigma_T := 4 p(1-p)$, where $p$ is the expectation of $p_i$.
The between-laboratory variance is defined as
\begin{equation*}
  \sigma_{L;O}^2:= \frac{4}{L} \sum_{i=1}^L (p_i - p)^2.
\end{equation*}
Since Gadrich and Bashkansky proved the following relationship among the variances above,
\begin{equation*}
  \sigma_T^2 = \frac{1}{L} \sum_{i=1}^L \sigma^2_{ri;O} + \sigma^2_{L;O},
\end{equation*}
the repeatability and reproducibility variances are respectively defined as
\begin{equation*}
  \sigma^2_{r;O} := \frac{1}{L} \sum_{i=1}^L \sigma^2_{ri;O}, \text{ and } \sigma^2_{R;O} := \sigma^2_T.
\end{equation*}

To estimate the repeatability, between-laboratory, and reproducibility variances, $p_i$ and $p$ are replaced with their unbiased variances.
Thus, Gadrich and Bashkansky proposed the following estimators:
\begin{align*}
  \hat{p}_i            & = \frac{1}{n} \sum_{j=1}^n y_{ij}, \quad \hat{p} = \frac{1}{L} \sum_{i=1}^L \hat{p}_i, \\%\label{eq:5725_est_p}\\
  \hat{\sigma}_{r;O}^2 & = \frac{4}{L} \sum_{i=1}^L \hat{p}_i (1 - \hat{p_i}),                                  \\%\label{eq:5725_est_sigmar}\\
  \hat{\sigma}_{L;O}^2 & = \frac{4}{L} \sum_{i=1}^L (\hat{p}_i - \hat{p})^2,\text{ and }                        \\
  %	\label{eq:5725_est_sigmaL}\\
  \hat{\sigma}_{R;O}^2 & = 4 \hat{p} (1-\hat{p}).
  %	\label{eq:unbiased_est_sigmaR}
\end{align*}

\section{Existing test methods for homogeneity of proportions}
\label{sec:test}

Throughout this section, let $\alpha \in (0,1)$.

\subsection{Standard chi-squared test}
\label{sec:test_std}

The statistic for the standard chi-squared test is defined by
\begin{equation}
  \label{eq:stat_std}
  I_{(S)} = \sum_{i=1}^L \frac{n \left(\hat{p}_i - \hat{p}\right)^2}{\hat{p} (1-\hat{p})},
\end{equation}
and to determine whether $H_0$ is rejected, the method uses the fact that $I_{(S)}$ approximately follows the chi-squared distribution with $(L-1)$-degree of freedom under $H_0$.
In other words, $H_0$ is rejected at the $\alpha$ significance level if $I_{(S)} > \chi_{L-1;\alpha}^2$, where $\chi_{L-1,\alpha}^2$ represents the upper $\alpha$ critical value of a chi-squared distribution with $(L-1)$ degrees of freedom.

Here we note that, in their models, \citet{wilrich2010DeterminationPrecisionQualitativeMeasurement} and \citet{gadrich2012ORDANOVAAnalysisOrdinalVariation} proposed applying the standard chi-squared test to detect laboratory effects.

\subsection{Potthoff and Whittinghill's test}
\label{sec:potth-whitt-test}
When $p$ is assumed to be known, the statistic of the Potthoff and Whittinghill test is defined by
\begin{equation*}
  I_{(PW)} = \frac{\sum_{i=1}^L x_i (x_i - 1)}{p}
  + \frac{\sum_{i=1}^L (n-x_i) (n - x_i - 1)}{1-p};
\end{equation*}
and to determine whether $H_0$ is rejected, the method uses the fact that $c_1(p) I_{(PW)} + c_2 (p)$ approximately follows $\nu (p)$ degrees of freedom under $H_0$, where $c_1(p)$, $c_2(p)$, and $\nu (p)$ are constants depending on $p$.
These constants  are chosen such that the mean, variance, and third central moment of $c_1 (p) I_{(PW)}+c_2(p)$ equal the mean, variances, and third central moment of the chi-squared distribution with $\nu (p)$ degrees of freedom, respectively.
In our environment, $c_1 (p)$, $c_2(p)$, and $\nu(p)$ are expressed as follows:
\begin{align*}
  c_1 (p) & = \frac{2}{p(1-p)-2 (n-4)},        \\
  c_2(p)  & = c_1 (p) (c_1 (p) - 1) L n (n-1), \\
  \nu (p) & = c_1 (p)^2 Ln(n-1).
\end{align*}

To handle the cases where $p$ is unknown, Potthoff and Whittinghill proposed that $p$ be set so as to minimizing $I_{(PW)}$.
The resulting values are
\begin{align*}
  p_{\min}       & = \frac{\left(\sum_{i=1}^L x_i (x_i-1)\right)^{1/2}}{\left(\sum_{i=1}^L x_i (x_i-1)\right)^{1/2} + \left(\sum_{i=1}^L (n-x_i) (n-x_i-1)\right)^{1/2}}, \\
  \label{eq:stat_PW_min}
  I_{(PW), \min} & = \left[ \left(\sum_{i=1}^L x_i (x_i-1)\right)^{1/2} + \left(\sum_{i=1}^L (n-x_i) (n-x_i-1)\right)^{1/2}\right]^2.
\end{align*}

In summary, $H_0$ is rejected with an $\alpha$ significance level if
\begin{equation*}
  c_1(p_{\text{min}}) I_{(PW),\text{min}} + c_2\left(p_{\text{min}}\right)>\chi_{\nu (p_{\text{min}}); \alpha}^2,
\end{equation*}
where $\chi_{\nu(p_{\text{min}}),\alpha}^2$ represents the upper $\alpha$ critical value of the chi-squared distribution with $\nu(p_{min})$ degrees of freedom.

\subsection{Nass's test method}
\label{sec:nasss-thest}

Nass's test method modifies the standard chi-squared test to improve its approximation for sparse data.
The method uses the same statistic as the standard chi-squared test method, but to determine whether $H_0$ is rejected, the method uses the fact that $c(\hat{p}) I_{(S)}$ is approximated by the $\nu (\hat{p})$ degrees of freedom under $H_0$, where $c(\hat{p})$ and $\nu(\hat{p})$ are constants depending on $\hat{p}$.
These constants are chosen such that the mean and variance of $c (\hat{p})I_{(S)}$ equal the mean and variance of the chi-squared distribution with $\nu (\hat{p})$ degrees of freedom, respectively.
In our environments, $c(\hat{p})$ and $\nu (\hat{p})$  are expressed as follows:
\begin{equation*}
  c(\hat{p}) = \frac{(Ln-3)(Ln-2)(Ln-1) \hat{p} (1- \hat{p})}{L (n-1) \left( L^2 n^2  \hat{p} (1-\hat{p}) - L n +1\right)}
\end{equation*}
and
\begin{equation*}
  \nu(\hat{p})= \frac{(Ln-3)(Ln-2) n (L-1) \hat{p} (1- \hat{p})}{ (n-1) \left( L^2 n^2  \hat{p} (1-\hat{p}) - L n +1\right)}. 
\end{equation*}
Here, $H_0$ is rejected with $\alpha$ significance level if $c(\hat{p}) I_{(S)} > \chi_{\nu (\hat{p});\alpha}^2$, where $\chi_{\nu,\alpha}^2$ represents the upper $\alpha$ critical value of a chi-squared distribution with $\nu(\hat{p})$ degrees of freedom.

If $\hat{p}=0$ or $1$, then $c(\hat{p}) = \nu(\hat{p}) = 0$; and if $\hat{p}= 1 / (Ln)$ or $(Ln-1) / (Ln)$, then $c(\hat{p}) =\nu(\hat{p}) = \infty$. 
In these cases, Nass's test procedure does not work well. 
For the case where $\hat{p}=0$ or $1$, we should consider that there are no laboratory effects, because all the laboratory's PODs $\hat{p}_i$ are completely identical; that is, $\hat{p}_i=0$ or $1$ for any $i \in \{1,\ldots,L\}$. 
On the other hand, the case where $\hat{p} = 1/(Ln)$ or $(Ln-1) / (Ln)$ means that there exists $i^\ast \in \{1,\ldots,L\}$ such that $x_{i^\ast}=1$ and $x_i=0 \; (i\neq i^\ast)$, or $x_{i^\ast}=0$ and $x_i = 1 \; (i \neq i^\ast)$. 
For the case where $x_{i^\ast}=1$, by considering $x_{i^\ast}=1+\varepsilon \; (\varepsilon>0)$ and observing the asymptotic behaviors of $c (\hat{p}) I_{(S)}|_\varepsilon$ and $\chi_{\nu(\hat{p}); \alpha}^2|_\varepsilon$ when $\varepsilon\rightarrow 0$, we can determine the Nass test result. 
Note that $c(\hat{p}) I_{(S)}|_\varepsilon$ and $\chi_{\nu(\hat{p});\alpha}^2|_\varepsilon$ express the values of $c(\hat{p}) I_{(S)}|_\varepsilon$ and $\chi_{\nu(\hat{p});\alpha}^2|_\varepsilon$  when $x_{i^\ast}=1+\varepsilon$. 
If $\varepsilon'$ is sufficiently large, the Nass test rejects $H_0$; that is, $c(\hat{p})I_{(S)}|_{\varepsilon'} >\chi_{\nu(\hat{p});\alpha}^2|_{\varepsilon'}$. 
Thus, if there exits $\varepsilon'>0$ such that
\begin{equation}
  \label{eq:Nass_asym}
  \begin{cases}
    c(\hat{p}) I_{(S)}|_{\varepsilon} > \chi^2_{\nu(\hat{p}); \alpha}|_{\varepsilon} & \text{if } \varepsilon > \varepsilon'   \\
    c(\hat{p}) I_{(S)}|_{\varepsilon} < \chi^2_{\nu(\hat{p}); \alpha}|_{\varepsilon} & \text{if } \varepsilon < \varepsilon',
  \end{cases}
\end{equation}
then $H_0$ should not be rejected; otherwise, $H_0$ should be rejected.
For the case where $x_{i^*} = n-1$, we should consider $x_{i^*} = n - 1 - \varepsilon \ (\varepsilon > 0)$ instead of $x_{i^*} = 1 + \varepsilon \ (\varepsilon > 0)$ in the above procedure.

\begin{figure}
  \begin{center}
    \includegraphics[scale=0.4]{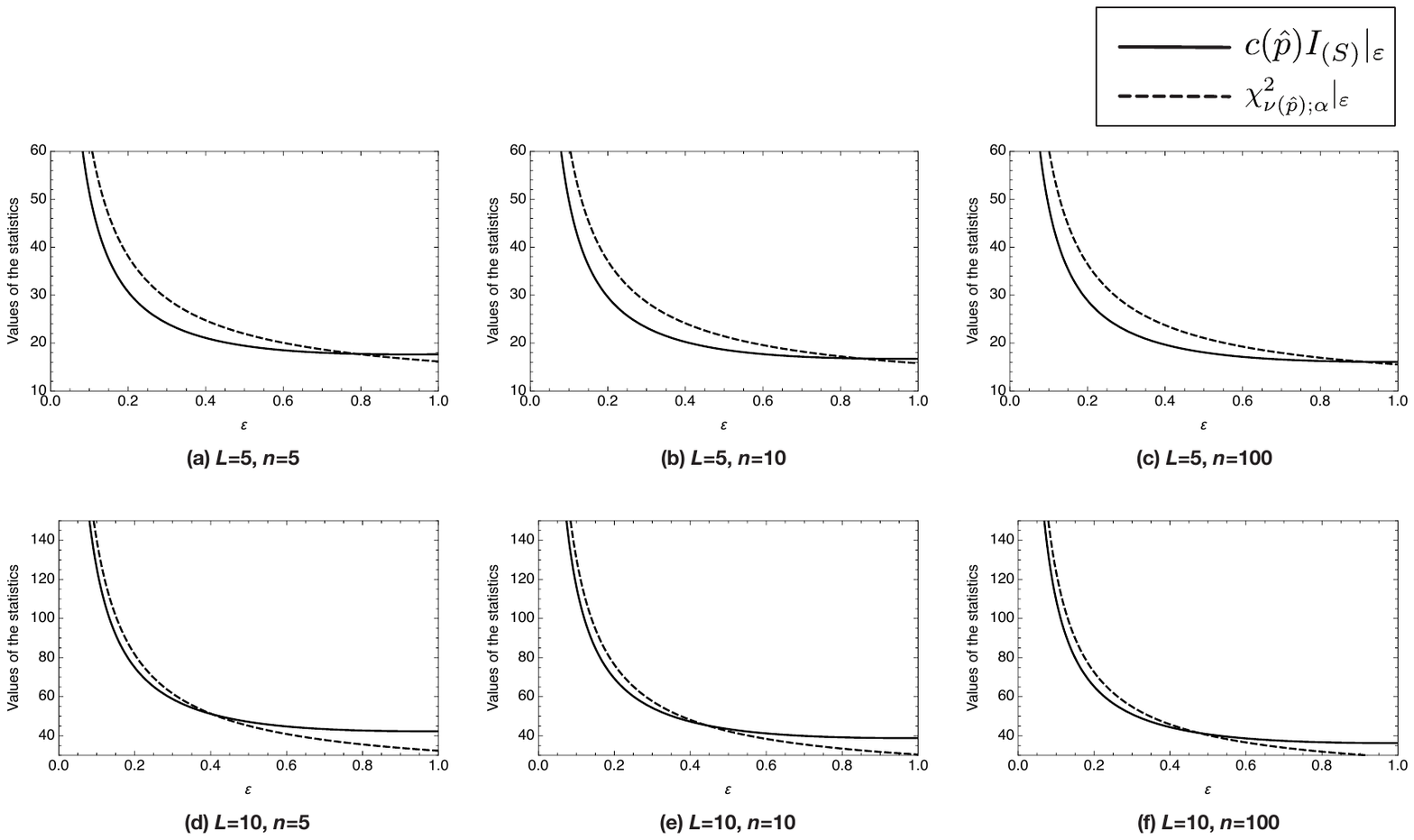}
  \end{center}
  \caption{
  Comparison between the Nass test statistic $c(\hat{p}) I_{(S)}|_\varepsilon$ and the upper $0.05$ critical value of a chi-squared distribution with $\nu(\hat{p})$ degrees of freedom $\chi_{\nu(\hat{p});0.05}^2|_\varepsilon$, when $\hat{p}=(1+\varepsilon) / (Ln)$, for the cases where $L = 5, 10$ and $n = 5, 10, 100$. 
  In each subfigure, the $x$-axis and $y$-axis are $\varepsilon$ and the value of the two statistics, respectively, and the solid and dashed lines show $c(\hat{p}) I_{(S)}|_\varepsilon$ and  $\chi^2_{\nu{(\hat{p})};0.05}|_{\varepsilon}$, respectively. 
  In all the cases, $c(\hat{p}) I_{(S)}|_\varepsilon\ <  \chi_{\nu(\hat{p});0.05}^2|_\varepsilon$ if $\varepsilon \approx 1$, and $c(\hat{p}) I_{(S)}|_\varepsilon\ > \chi_{\nu(\hat{p});0.05}^2|_\varepsilon$ if $\varepsilon \approx 0$.
  \label{fig1}}
\end{figure}

For example, consider the cases where the number of laboratories $L$ is set at $5$ and $10$, the number of measured values in each laboratory $n$ is set at $5$, $10$, and $100,$ and $x_{i^\ast}=1+\varepsilon$ and $x_i=0 \ (i\neq i^\ast)$. 
In all these cases, there exists $\varepsilon' > 0$ satisfying. 
In Figure~\ref{fig1}, the solid and dashed lines show $c(\hat{p})I_{(S)}|_\varepsilon$ and $\chi_{\nu(\hat{p});\alpha}^2|_\varepsilon$, respectively, for the cases where $L = 5, 10$ and $n = 5, 10, 100$, in the range of $\varepsilon \in (0, 1)$. 
In each subfigure, we see that $c(\hat{p})I_{(S)}|_\varepsilon  < \chi_{\nu(\hat{p});\alpha}^2|_\varepsilon$  if $\varepsilon \approx 1$, and $c(\hat{p})I_{(S)}|_\varepsilon > \chi_{\nu(\hat{p});\alpha}^2|_\varepsilon$ if $\varepsilon \approx 0$. 
As both the $c(\hat{p})I_{(S)}|_\varepsilon$ and $\chi_{\nu(\hat{p});\alpha}^2|_\varepsilon$ curves are continuous, there exists $\varepsilon' > 0$ satisfying~\eqref{eq:Nass_asym}. 
Thus, $H_0$ should not be rejected when $\hat{p}=1/(Ln)$. 

\subsection{Xu's test method}
\label{sec:xus-test-method}

The test method proposed by Xu~\cite[Chpter 5.2]{xu2011StatisticalIssuesMetaAnalysis} is a relatively new procedure.
The statistic for the Xu test is given by
\begin{equation*}
  \label{eq:stat_Xu}
  I_{(Xu)} = \sqrt{\frac{n(n-1)}{2L}}\frac{1}{\hat{p} (1-\hat{p})} \sum_{i=1}^L U_i,
\end{equation*}
where $U_i:= \left(\hat{p}_i - \hat{p}\right)^2 - (L-1)/(L(n-1)) \hat{p}_i (1-\hat{p}_i)$.
Since, for $U_i$,
\begin{align*}
   & E(U_i) = 0,                                                                                  \\ 
   & \mathrm{Var} (U_i) = \frac{2 p^2 (1-p)^2}{n(n-1)} + O\left(\frac{1}{nL}\right), \text{ and } \\
   & \mathrm{Cov} (U_i, U_j) = O\left(\frac{1}{nL}\right) \text{ for } i \neq j.
\end{align*}
Xu proved that $I_{(Xu)}$ converges to the standard normal distribution as $n L \to \infty$ under $H_0$.
Thus, $H_0$ is rejected with an $\alpha>0$ significance level if $I_{(Xu)}>z_{1-\alpha}$, where $z_{1-\alpha}$ represents the upper $\alpha$ critical value of the standard normal distribution.

Note that, for the case where $\hat{p}= 0$ or $1$, we should consider that there are no laboratory effects, because all the laboratory's PODs $\hat{p}_i$ are completely identical.

\clearpage
\part*{Supplementary Tables}

\begin{sidewaystable}
  \tiny
  \centering
  \caption{\small The average, lower $2.5\%$-tile, and upper $2.5\%$-tile estimates of POD $p$, $\sigma_r^2$, $\sigma_L^2$, and $\sigma_R^2$ in the simulated study for $L=5$.
    In the table, $a$ and $b$ are the parameters of the assumed beta-distributions, $L$ and $n$ are the numbers of laboratories participating in the collaborative study and the repetitions of each laboratory, and $p$ and $\lambda$ are the mean and over-dispersion parameter of the assumed beta-distributions, respectively. 
    $\sigma_r^2$, $\sigma_L^2$, and $\sigma_R^2$ are the theoretical repeatability, between-laboratory, and reproducibility variances, respectively, of the assumed beta-binomial models, and the symbol $\hat{\cdot}$ means an estimate. 
    The subscript letters $E$, $l$, and $u$ represent the average, lower $2.5\%$-tile, and upper $2.5\%$-tile estimates, respectively.   \label{tab:estres_L5}}
  \begin{tabular}{rrrr|rr|rrr|rrrr|rrrr|rrrr}\hline
    $a$     & $b$    & $L$ & $n$   & $p$    & $\lambda$ & $\hat{p}_E$ & $\hat{p}_l$ & $\hat{p}_u $ & $\sigma^2_r$ & $\hat{\sigma}^2_{r,E}$ & $\hat{\sigma}^2_{r,l}$ & $\hat{\sigma}^2_{r,u}$ & $\sigma^2_L$ & $\hat{\sigma}^2_{L,E}$ & $\sigma^2_{L,l}$ & $\hat{\sigma}^2_{L,u}$ & $\sigma^2_R$ & $\hat{\sigma}^2_{R,E}$ & $\sigma^2_{R,l}$ & $\hat{\sigma}^2_{R,u}$ \\\hline
    $13.3$  & $5.7$  & $5$ & $5$   & $0.7$  & $0.05$    & $0.701$     & $0.480$     & $0.880$      & $0.200$      & $0.199$                & $0.100$                & $0.280$                & $0.011$      & $0.011$                & $-0.040$         & $0.108$                & $0.210$      & $0.210$                & $0.108$          & $0.268$                \\
    $13.3$  & $5.7$  & $5$ & $10$  & $0.7$  & $0.05$    & $0.702$     & $0.540$     & $0.840$      & $0.200$      & $0.199$                & $0.124$                & $0.256$                & $0.011$      & $0.010$                & $-0.018$         & $0.065$                & $0.210$      & $0.209$                & $0.136$          & $0.256$                \\
    $13.3$  & $5.7$  & $5$ & $100$ & $0.7$  & $0.05$    & $0.701$     & $0.602$     & $0.794$      & $0.200$      & $0.199$                & $0.157$                & $0.233$                & $0.011$      & $0.010$                & $-0.000$         & $0.033$                & $0.210$      & $0.210$                & $0.165$          & $0.243$                \\
    $6.3$   & $2.7$  & $5$ & $5$   & $0.7$  & $0.1$     & $0.701$     & $0.480$     & $0.880$      & $0.189$      & $0.189$                & $0.080$                & $0.280$                & $0.021$      & $0.021$                & $-0.040$         & $0.132$                & $0.210$      & $0.210$                & $0.108$          & $0.272$                \\
    $6.3$   & $2.7$  & $5$ & $10$  & $0.7$  & $0.1$     & $0.701$     & $0.520$     & $0.860$      & $0.189$      & $0.188$                & $0.107$                & $0.253$                & $0.021$      & $0.021$                & $-0.017$         & $0.092$                & $0.210$      & $0.210$                & $0.122$          & $0.263$                \\
    $6.3$   & $2.7$  & $5$ & $100$ & $0.7$  & $0.1$     & $0.701$     & $0.568$     & $0.824$      & $0.189$      & $0.189$                & $0.134$                & $0.233$                & $0.021$      & $0.021$                & $0.001$          & $0.060$                & $0.210$      & $0.210$                & $0.147$          & $0.252$                \\
    $0.7$   & $0.3$  & $5$ & $5$   & $0.7$  & $0.5$     & $0.701$     & $0.360$     & $0.960$      & $0.105$      & $0.105$                & $0.000$                & $0.220$                & $0.105$      & $0.105$                & $-0.004$         & $0.260$                & $0.210$      & $0.209$                & $0.040$          & $0.300$                \\
    $0.7$   & $0.3$  & $5$ & $10$  & $0.7$  & $0.5$     & $0.699$     & $0.380$     & $0.960$      & $0.105$      & $0.105$                & $0.020$                & $0.202$                & $0.105$      & $0.105$                & $-0.000$         & $0.244$                & $0.210$      & $0.210$                & $0.039$          & $0.294$                \\
    $0.7$   & $0.3$  & $5$ & $100$ & $0.7$  & $0.5$     & $0.700$     & $0.382$     & $0.950$      & $0.105$      & $0.105$                & $0.028$                & $0.189$                & $0.105$      & $0.104$                & $0.003$          & $0.226$                & $0.210$      & $0.210$                & $0.048$          & $0.290$                \\
    $17.1$  & $1.9$  & $5$ & $5$   & $0.9$  & $0.05$    & $0.900$     & $0.760$     & $1.000$      & $0.086$      & $0.085$                & $0.000$                & $0.200$                & $0.005$      & $0.005$                & $-0.016$         & $0.056$                & $0.090$      & $0.090$                & $0.000$          & $0.196$                \\
    $17.1$  & $1.9$  & $5$ & $10$  & $0.9$  & $0.05$    & $0.900$     & $0.780$     & $0.980$      & $0.086$      & $0.085$                & $0.020$                & $0.164$                & $0.005$      & $0.004$                & $-0.010$         & $0.033$                & $0.090$      & $0.090$                & $0.020$          & $0.175$                \\
    $17.1$  & $1.9$  & $5$ & $100$ & $0.9$  & $0.05$    & $0.900$     & $0.830$     & $0.956$      & $0.086$      & $0.085$                & $0.042$                & $0.135$                & $0.005$      & $0.004$                & $-0.000$         & $0.017$                & $0.090$      & $0.090$                & $0.042$          & $0.144$                \\
    $8.1$   & $0.9$  & $5$ & $5$   & $0.9$  & $0.1$     & $0.900$     & $0.720$     & $1.000$      & $0.081$      & $0.081$                & $0.000$                & $0.180$                & $0.009$      & $0.009$                & $-0.016$         & $0.084$                & $0.090$      & $0.090$                & $0.000$          & $0.208$                \\
    $8.1$   & $0.9$  & $5$ & $10$  & $0.9$  & $0.1$     & $0.901$     & $0.760$     & $1.000$      & $0.081$      & $0.081$                & $0.000$                & $0.164$                & $0.009$      & $0.009$                & $-0.006$         & $0.054$                & $0.090$      & $0.090$                & $0.000$          & $0.187$                \\
    $8.1$   & $0.9$  & $5$ & $100$ & $0.9$  & $0.1$     & $0.900$     & $0.802$     & $0.970$      & $0.081$      & $0.081$                & $0.028$                & $0.141$                & $0.009$      & $0.009$                & $0.000$          & $0.037$                & $0.090$      & $0.090$                & $0.029$          & $0.164$                \\
    $0.9$   & $0.1$  & $5$ & $5$   & $0.9$  & $0.5$     & $0.899$     & $0.640$     & $1.000$      & $0.045$      & $0.045$                & $0.000$                & $0.140$                & $0.045$      & $0.045$                & $-0.004$         & $0.200$                & $0.090$      & $0.091$                & $0.000$          & $0.268$                \\
    $0.9$   & $0.1$  & $5$ & $10$  & $0.9$  & $0.5$     & $0.898$     & $0.640$     & $1.000$      & $0.045$      & $0.045$                & $0.000$                & $0.136$                & $0.045$      & $0.046$                & $-0.001$         & $0.200$                & $0.090$      & $0.091$                & $0.000$          & $0.265$                \\
    $0.9$   & $0.1$  & $5$ & $100$ & $0.9$  & $0.5$     & $0.898$     & $0.654$     & $1.000$      & $0.045$      & $0.045$                & $0.000$                & $0.126$                & $0.045$      & $0.046$                & $0.000$          & $0.188$                & $0.090$      & $0.091$                & $0.000$          & $0.257$                \\
    $18.05$ & $0.95$ & $5$ & $5$   & $0.95$ & $0.05$    & $0.950$     & $0.840$     & $1.000$      & $0.045$      & $0.045$                & $0.000$                & $0.140$                & $0.002$      & $0.002$                & $-0.012$         & $0.024$                & $0.048$      & $0.047$                & $0.000$          & $0.148$                \\
    $18.05$ & $0.95$ & $5$ & $10$  & $0.95$ & $0.05$    & $0.950$     & $0.860$     & $1.000$      & $0.045$      & $0.045$                & $0.000$                & $0.111$                & $0.002$      & $0.002$                & $-0.004$         & $0.023$                & $0.048$      & $0.047$                & $0.000$          & $0.123$                \\
    $18.05$ & $0.95$ & $5$ & $100$ & $0.95$ & $0.05$    & $0.950$     & $0.898$     & $0.986$      & $0.045$      & $0.045$                & $0.014$                & $0.087$                & $0.002$      & $0.002$                & $-0.000$         & $0.011$                & $0.048$      & $0.047$                & $0.014$          & $0.094$                \\
    $8.55$  & $0.45$ & $5$ & $5$   & $0.95$ & $0.1$     & $0.950$     & $0.840$     & $1.000$      & $0.043$      & $0.043$                & $0.000$                & $0.140$                & $0.005$      & $0.005$                & $-0.012$         & $0.060$                & $0.048$      & $0.047$                & $0.000$          & $0.148$                \\
    $8.55$  & $0.45$ & $5$ & $10$  & $0.95$ & $0.1$     & $0.950$     & $0.840$     & $1.000$      & $0.043$      & $0.043$                & $0.000$                & $0.114$                & $0.005$      & $0.005$                & $-0.003$         & $0.039$                & $0.048$      & $0.047$                & $0.000$          & $0.139$                \\
    $8.55$  & $0.45$ & $5$ & $100$ & $0.95$ & $0.1$     & $0.950$     & $0.872$     & $0.994$      & $0.043$      & $0.043$                & $0.006$                & $0.097$                & $0.005$      & $0.005$                & $0.000$          & $0.025$                & $0.048$      & $0.047$                & $0.006$          & $0.116$                \\
    $0.95$  & $0.05$ & $5$ & $5$   & $0.95$ & $0.5$     & $0.949$     & $0.760$     & $1.000$      & $0.024$      & $0.024$                & $0.000$                & $0.100$                & $0.024$      & $0.024$                & $0.000$          & $0.200$                & $0.048$      & $0.049$                & $0.000$          & $0.220$                \\
    $0.95$  & $0.05$ & $5$ & $10$  & $0.95$ & $0.5$     & $0.949$     & $0.760$     & $1.000$      & $0.024$      & $0.024$                & $0.000$                & $0.100$                & $0.024$      & $0.025$                & $0.000$          & $0.190$                & $0.048$      & $0.049$                & $0.000$          & $0.213$                \\
    $0.95$  & $0.05$ & $5$ & $100$ & $0.95$ & $0.5$     & $0.949$     & $0.764$     & $1.000$      & $0.024$      & $0.024$                & $0.000$                & $0.093$                & $0.024$      & $0.024$                & $0.000$          & $0.169$                & $0.048$      & $0.049$                & $0.000$          & $0.206$                \\\bottomrule
  \end{tabular}
\end{sidewaystable}

\begin{sidewaystable}
  \centering
  \tiny
  \caption{\small The average, lower $2.5\%$-tile, and upper $2.5\%$-tile estimates of POD $p$, $\sigma_r^2$, $\sigma_L^2$, and $\sigma_R^2$ in the simulated study for $L=10$.
    In the table, $a$ and $b$ are the parameters of the assumed beta-distributions, $L$ and $n$ are the numbers of laboratories participating in the collaborative study and the repetitions of each laboratory, and $p$ and $\lambda$ are the mean and over-dispersion parameter of the assumed beta-distributions, respectively. 
    $\sigma_r^2$, $\sigma_L^2$, and $\sigma_R^2$ are the theoretical repeatability, between-laboratory, and reproducibility variances, respectively, of the assumed beta-binomial models, and the symbol $\hat{\cdot}$ means an estimate. 
    The subscript letters $E$, $l$, and $u$ represent the average, lower $2.5\%$-tile, and upper $2.5\%$-tile estimates, respectively.  \label{tab:estres_L10}}
  \begin{tabular}{rrrr|rr|rrr|rrrr|rrrr|rrrr}\hline
    $a$     & $b$    & $L$  & $n$   & $p$    & $\lambda$ & $\hat{p}_E$ & $\hat{p}_l$ & $\hat{p}_u $ & $\sigma^2_r$ & $\hat{\sigma}^2_{r,E}$ & $\hat{\sigma}^2_{r,l}$ & $\hat{\sigma}^2_{r,u}$ & $\sigma^2_L$ & $\hat{\sigma}^2_{L,E}$ & $\sigma^2_{L,l}$ & $\hat{\sigma}^2_{L,u}$ & $\sigma^2_R$ & $\hat{\sigma}^2_{R,E}$ & $\sigma^2_{R,l}$ & $\hat{\sigma}^2_{R,u}$ \\\hline
    $13.3$  & $5.7$  & $10$ & $5$   & $0.7$  & $0.05$    & $0.700$     & $0.560$     & $0.840$      & $0.200$      & $0.200$                & $0.130$                & $0.260$                & $0.011$      & $0.010$                & $-0.030$         & $0.071$                & $0.210$      & $0.210$                & $0.139$          & $0.253$                \\
    $13.3$  & $5.7$  & $10$ & $10$  & $0.7$  & $0.05$    & $0.700$     & $0.590$     & $0.800$      & $0.200$      & $0.200$                & $0.147$                & $0.242$                & $0.011$      & $0.011$                & $-0.012$         & $0.045$                & $0.210$      & $0.210$                & $0.161$          & $0.245$                \\
    $13.3$  & $5.7$  & $10$ & $100$ & $0.7$  & $0.05$    & $0.700$     & $0.630$     & $0.767$      & $0.200$      & $0.200$                & $0.171$                & $0.224$                & $0.011$      & $0.011$                & $0.002$          & $0.025$                & $0.210$      & $0.210$                & $0.180$          & $0.235$                \\
    $6.3$   & $2.7$  & $10$ & $5$   & $0.7$  & $0.1$     & $0.700$     & $0.540$     & $0.840$      & $0.189$      & $0.189$                & $0.110$                & $0.260$                & $0.021$      & $0.021$                & $-0.026$         & $0.089$                & $0.210$      & $0.210$                & $0.137$          & $0.256$                \\
    $6.3$   & $2.7$  & $10$ & $10$  & $0.7$  & $0.1$     & $0.700$     & $0.570$     & $0.820$      & $0.189$      & $0.189$                & $0.133$                & $0.237$                & $0.021$      & $0.021$                & $-0.008$         & $0.064$                & $0.210$      & $0.210$                & $0.151$          & $0.250$                \\
    $6.3$   & $2.7$  & $10$ & $100$ & $0.7$  & $0.1$     & $0.700$     & $0.603$     & $0.788$      & $0.189$      & $0.189$                & $0.152$                & $0.221$                & $0.021$      & $0.021$                & $0.005$          & $0.046$                & $0.210$      & $0.210$                & $0.169$          & $0.242$                \\
    $0.7$   & $0.3$  & $10$ & $5$   & $0.7$  & $0.5$     & $0.700$     & $0.460$     & $0.900$      & $0.105$      & $0.105$                & $0.030$                & $0.190$                & $0.105$      & $0.105$                & $0.006$          & $0.202$                & $0.210$      & $0.210$                & $0.092$          & $0.268$                \\
    $0.7$   & $0.3$  & $10$ & $10$  & $0.7$  & $0.5$     & $0.700$     & $0.480$     & $0.900$      & $0.105$      & $0.105$                & $0.041$                & $0.174$                & $0.105$      & $0.105$                & $0.016$          & $0.190$                & $0.210$      & $0.210$                & $0.095$          & $0.266$                \\
    $0.7$   & $0.3$  & $10$ & $100$ & $0.7$  & $0.5$     & $0.700$     & $0.486$     & $0.890$      & $0.105$      & $0.105$                & $0.049$                & $0.164$                & $0.105$      & $0.105$                & $0.023$          & $0.181$                & $0.210$      & $0.210$                & $0.101$          & $0.265$                \\
    $17.1$  & $1.9$  & $10$ & $5$   & $0.9$  & $0.05$    & $0.900$     & $0.800$     & $0.980$      & $0.086$      & $0.086$                & $0.020$                & $0.160$                & $0.005$      & $0.004$                & $-0.013$         & $0.036$                & $0.090$      & $0.090$                & $0.020$          & $0.164$                \\
    $17.1$  & $1.9$  & $10$ & $10$  & $0.9$  & $0.05$    & $0.900$     & $0.820$     & $0.960$      & $0.086$      & $0.086$                & $0.038$                & $0.140$                & $0.005$      & $0.004$                & $-0.005$         & $0.024$                & $0.090$      & $0.090$                & $0.039$          & $0.149$                \\
    $17.1$  & $1.9$  & $10$ & $100$ & $0.9$  & $0.05$    & $0.900$     & $0.851$     & $0.940$      & $0.086$      & $0.086$                & $0.055$                & $0.121$                & $0.005$      & $0.004$                & $0.001$          & $0.013$                & $0.090$      & $0.090$                & $0.057$          & $0.128$                \\
    $8.1$   & $0.9$  & $10$ & $5$   & $0.9$  & $0.1$     & $0.900$     & $0.780$     & $0.980$      & $0.081$      & $0.081$                & $0.020$                & $0.160$                & $0.009$      & $0.009$                & $-0.009$         & $0.055$                & $0.090$      & $0.090$                & $0.020$          & $0.175$                \\
    $8.1$   & $0.9$  & $10$ & $10$  & $0.9$  & $0.1$     & $0.900$     & $0.810$     & $0.970$      & $0.081$      & $0.081$                & $0.028$                & $0.139$                & $0.009$      & $0.009$                & $-0.003$         & $0.039$                & $0.090$      & $0.090$                & $0.029$          & $0.157$                \\
    $8.1$   & $0.9$  & $10$ & $100$ & $0.9$  & $0.1$     & $0.900$     & $0.833$     & $0.952$      & $0.081$      & $0.081$                & $0.043$                & $0.123$                & $0.009$      & $0.009$                & $0.001$          & $0.026$                & $0.090$      & $0.090$                & $0.046$          & $0.141$                \\
    $0.9$   & $0.1$  & $10$ & $5$   & $0.9$  & $0.5$     & $0.899$     & $0.720$     & $1.000$      & $0.045$      & $0.045$                & $0.000$                & $0.110$                & $0.045$      & $0.045$                & $-0.001$         & $0.158$                & $0.090$      & $0.090$                & $0.000$          & $0.212$                \\
    $0.9$   & $0.1$  & $10$ & $10$  & $0.9$  & $0.5$     & $0.899$     & $0.740$     & $1.000$      & $0.045$      & $0.045$                & $0.000$                & $0.106$                & $0.045$      & $0.045$                & $-0.000$         & $0.146$                & $0.090$      & $0.090$                & $0.000$          & $0.207$                \\
    $0.9$   & $0.1$  & $10$ & $100$ & $0.9$  & $0.5$     & $0.900$     & $0.744$     & $0.996$      & $0.045$      & $0.045$                & $0.004$                & $0.099$                & $0.045$      & $0.045$                & $0.000$          & $0.139$                & $0.090$      & $0.090$                & $0.004$          & $0.202$                \\
    $18.05$ & $0.95$ & $10$ & $5$   & $0.95$ & $0.05$    & $0.950$     & $0.880$     & $1.000$      & $0.045$      & $0.045$                & $0.000$                & $0.110$                & $0.002$      & $0.002$                & $-0.005$         & $0.027$                & $0.048$      & $0.048$                & $0.000$          & $0.110$                \\
    $18.05$ & $0.95$ & $10$ & $10$  & $0.95$ & $0.05$    & $0.950$     & $0.890$     & $0.990$      & $0.045$      & $0.045$                & $0.010$                & $0.092$                & $0.002$      & $0.002$                & $-0.002$         & $0.015$                & $0.048$      & $0.048$                & $0.010$          & $0.100$                \\
    $18.05$ & $0.95$ & $10$ & $100$ & $0.95$ & $0.05$    & $0.950$     & $0.914$     & $0.978$      & $0.045$      & $0.045$                & $0.021$                & $0.074$                & $0.002$      & $0.002$                & $0.000$          & $0.008$                & $0.048$      & $0.048$                & $0.022$          & $0.079$                \\
    $8.55$  & $0.45$ & $10$ & $5$   & $0.95$ & $0.1$     & $0.950$     & $0.860$     & $1.000$      & $0.043$      & $0.042$                & $0.000$                & $0.100$                & $0.005$      & $0.005$                & $-0.005$         & $0.035$                & $0.048$      & $0.047$                & $0.000$          & $0.123$                \\
    $8.55$  & $0.45$ & $10$ & $10$  & $0.95$ & $0.1$     & $0.950$     & $0.880$     & $1.000$      & $0.043$      & $0.043$                & $0.000$                & $0.091$                & $0.005$      & $0.005$                & $-0.001$         & $0.029$                & $0.048$      & $0.047$                & $0.000$          & $0.108$                \\
    $8.55$  & $0.45$ & $10$ & $100$ & $0.95$ & $0.1$     & $0.950$     & $0.899$     & $0.985$      & $0.043$      & $0.043$                & $0.014$                & $0.080$                & $0.005$      & $0.005$                & $0.000$          & $0.018$                & $0.048$      & $0.047$                & $0.015$          & $0.092$                \\
    $0.95$  & $0.05$ & $10$ & $5$   & $0.95$ & $0.5$     & $0.950$     & $0.820$     & $1.000$      & $0.024$      & $0.024$                & $0.000$                & $0.080$                & $0.024$      & $0.024$                & $-0.001$         & $0.117$                & $0.048$      & $0.048$                & $0.000$          & $0.162$                \\
    $0.95$  & $0.05$ & $10$ & $10$  & $0.95$ & $0.5$     & $0.949$     & $0.820$     & $1.000$      & $0.024$      & $0.024$                & $0.000$                & $0.073$                & $0.024$      & $0.024$                & $-0.000$         & $0.109$                & $0.048$      & $0.048$                & $0.000$          & $0.158$                \\
    $0.95$  & $0.05$ & $10$ & $100$ & $0.95$ & $0.5$     & $0.950$     & $0.826$     & $1.000$      & $0.024$      & $0.024$                & $0.000$                & $0.069$                & $0.024$      & $0.024$                & $0.000$          & $0.103$                & $0.048$      & $0.048$                & $0.000$          & $0.154$                \\\bottomrule
  \end{tabular}
\end{sidewaystable}

%\end{appendices}
%%===========================================================================================%%
%% If you are submitting to one of the Nature Portfolio journals, using the eJP submission   %%
%% system, please include the references within the manuscript file itself. You may do this  %%
%% by copying the reference list from your .bbl file, paste it into the main manuscript .tex %%
%% file, and delete the associated \verb+\bibliography+ commands.                            %%
%%===========================================================================================%%

\clearpage

\renewcommand{\sc}[1]{\textsc{#1}}
\bibliographystyle{siam}
\bibliography{PrecisionForBinary}
%% if required, the content of .bbl file can be included here once bbl is generated
%%\input sn-article.bbl

\end{document}